\documentclass{article}

\usepackage{arxiv}
\usepackage[utf8]{inputenc} 
\usepackage[T1]{fontenc}    
\usepackage{amsthm,amsmath,amsfonts,amssymb}
\usepackage{hyperref}       
\usepackage{url}            
\usepackage{booktabs}       
\usepackage{nicefrac}       
\usepackage{microtype}      
\usepackage{lipsum}		
\usepackage{graphicx}
\usepackage{cite}
\usepackage{mathabx}
\usepackage{doi}

\title{Neuro-oscillatory models of cortical speech processing}

\author{ 
\hspace{1mm}Olesia Dogonasheva \\
	  Université Paris Cité, \\
        Institut Pasteur, AP-HP, Inserm,\\ 
        Fondation Pour l'Audition, Institut de l’Audition, \\
        IHU reConnect, F-75012,\\
	Paris, France \\
\And
\hspace{1mm}Anne-Lise Giraud \\
        Université Paris Cité, \\
        Institut Pasteur, AP-HP, Inserm,\\ 
        Fondation Pour l'Audition, Institut de l’Audition, \\
        IHU reConnect, F-75012,\\
	Paris, France \\
\And
\hspace{1mm}Denis Zakharov \\
        Centre for Cognition and Decision Making\\
        HSE University\\
	Moscow, Russia \\
\And
\hspace{1mm}Boris Gutkin \\
        Group of Neural Theory,\\
        École Normale Supérieure PSL*\\
	Paris, France \\
}

\renewcommand{\shorttitle}{Review: Computational approaches to brain-based mechanisms of speech processing}

\hypersetup{
pdftitle={Computational models of speech processing},
pdfauthor={Olesia Dogonasheva, Anne-Lise Giraud, Boris Gutkin, Denis Zakharov},
pdfkeywords={Speech recognition, Computational modelling, Auditory cortex, Speech processing, Audio processing, Predictive coding, Interactive activation}
}

\begin{document}
\maketitle

\begin{abstract}
In this review, we examine computational models that explore the role of neural oscillations in speech perception, spanning from early auditory processing to higher cognitive stages. We focus on models that use rhythmic brain activities, such as gamma, theta, and delta oscillations, to encode phonemes, segment speech into syllables and words, and integrate linguistic elements to infer meaning. We analyze the mechanisms underlying these models, their biological plausibility, and their potential applications in processing and understanding speech in real time, a computational feature that is achieved by the human brain but not yet implemented in speech recognition models. Real-time processing enables dynamic adaptation to incoming speech, allowing systems to handle the rapid and continuous flow of auditory information required for effective communication, interactive applications, and accurate speech recognition in a variety of real-world settings. While significant progress has been made in modeling the neural basis of speech perception, challenges remain, particularly in accounting for the complexity of semantic processing and the integration of contextual influences. Moreover, the high computational demands of biologically realistic models pose practical difficulties for their implementation and analysis. Despite these limitations, these models provide valuable insights into the neural mechanisms of speech perception. We conclude by identifying current limitations, proposing future research directions, and suggesting how these models can be further developed to achieve a more comprehensive understanding of speech processing in the human brain.
\end{abstract}


\section*{Introduction}

Understanding how the human brain processes and comprehends spoken language is a complex challenge with significant implications across fundamental neuroscience, cognitive psychology, and speech recognition technologies. Speech perception involves the rapid interpretation of acoustic signals to extract the meaningful content of words and phrases in real-time. This intricate process relies on multiple interacting neural mechanisms that operate across different temporal and spatial scales.

\begin{figure}[h]
\begin{minipage}[h]{1\linewidth}
\center{\includegraphics[width=0.9\linewidth]{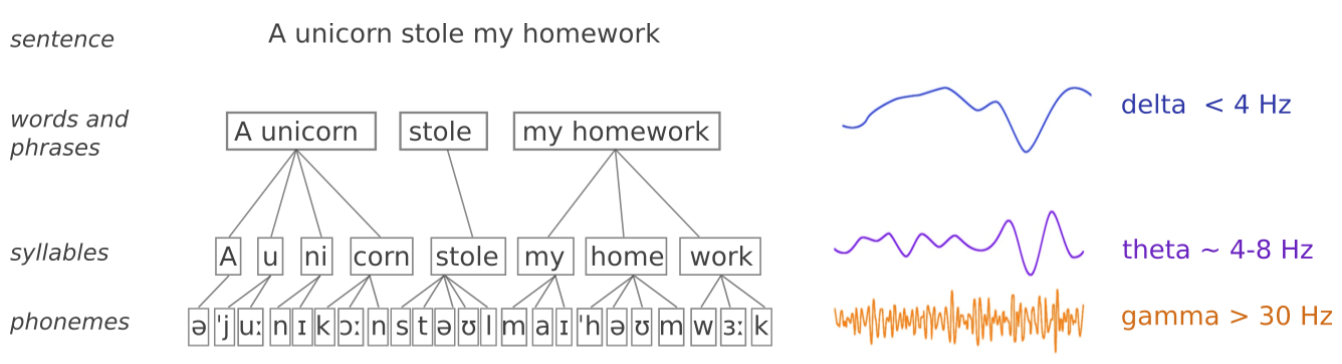}}
\end{minipage}
\caption{\textbf{Speech-coding hierarchy hypothesis.} The hypothesis suggests that speech is processed through a series of nested, hierarchical levels corresponding to different linguistic units, from phonemes to syllables, words, phrases, and sentences. At each level, specific neural oscillations—such as gamma, theta, and delta rhythms—are hypothesized to play a crucial role. Gamma rhythms (~30 Hz) are associated with the encoding of phonemic information, representing the smallest units of speech. Theta rhythms (~4–8 Hz) are proposed to support the segmentation of speech into syllables, aligning with the rhythm of syllabic units. Delta rhythms (below 4 Hz) are thought to facilitate the integration of longer linguistic elements such as words and phrases. The rhythmic synchronization is crucial for real-time speech recognition, allowing the brain to dynamically adapt to the temporal structure of spoken language.}
\label{fig:fig1}
\end{figure}

One prominent hypothesis suggests that the hierarchical rhythmic structure of speech—spanning phonemes, syllables, words, and sentences—provides a reliable basis for speech recognition in the brain (Fig.~\ref{fig:fig1}) \cite{stephenson2019untangling, greenberg1997modulation, kosem2016neural, kosem2018neural}. This hierarchical organization reflects various levels of linguistic processing, from the phonetic and morphological level, where phonemes combine into syllables and words, to the syntactic level, where words are arranged into meaningful phrases and sentences \cite{longacre1970hierarchy}. The neural processing of these hierarchical structures is thought to enhance the efficiency and accuracy of speech perception and production, a defining characteristic of human language behavior \cite{okada2010hierarchical, evans2015hierarchical, obleser2010segregation, de2017hierarchical, caucheteux2023evidence}.

On the other hand, brain rhythms represent plausible neural mechanisms to maintain a hierarchical structure during speech processing, as they map the speech structure \cite{chen2022ensemble, buzsaki2022brain, hyafil2015neural, poeppel2012maps, giraud2012cortical, ding2016cortical,ronconi2017multiple}. Because speech is a continuous signal, its parsing into linguistic elements of different granularity (phonemes, syllables, words, phrases) requires hierarchically organized real-time sampling and integration mechanisms \cite{ronconi2017multiple, gross2013speech}. This review aims to explore and synthesize the current state of the art of computational models that focus on the role of neural oscillations in speech perception. We analyze models that range from those implementing oscillatory dynamics in a biologically plausible manner to those employing more abstract approaches to study specific aspects of speech processing. We discuss the key questions these models address, the neural mechanisms they propose, the experiments used to validate their hypotheses, and their potential applications. Additionally, we examine the limitations of these models and highlight open questions and challenges for future research.

By synthesizing insights from a diverse set of computational models, this review seeks to advance our understanding of the neurocomputational principles underlying speech perception and to identify promising directions for both neuroscience and artificial intelligence.

\section{Rhythm-based models of speech processing}

After the initial processing of the auditory signal, where speech information is converted into a frequency spectrogram (cochlea), the auditory system performs further analyses in a highly hierarchical fashion. Contrary to the visual system, the auditory system has many relays such that the signal that reaches the auditory cortex is already preprocessed and a number of cues are already extracted, notably spatial information, auditory texture, alert nature of signals, etc. Many auditory tasks can be performed by animals without the auditory cortex, including sound discrimination \cite{abeles1975behavioral, kelly1986effects, heffner1990effect}. The role of the auditory is therefore largely dedicated to “sampling/parsing” and “integration/chunking” processes, which consist in sacrificing spectro-temporal details to interpret signals and put them in a multisensory and semantic context to plan appropriate responses. In humans, the left temporal cortex is dedicated to speech and language processing, while the right one is primarily involved in pitch, music, and prosody processing. This division of labor partly relies on different scales of temporal integration: fast and slow integration, respectively \cite{zatorre2002structure, poeppel2003analysis, neophytou2022differences}.

\textit{BOX HERE : Models of peripheral processing.}

The question of how the auditory cortex extracts meaningful speech units from the spectrograms remains a topic of active debate. On the shortest timescale, phonemes—the basic elements of spoken language—must be accurately recognized to understand spoken words and phrases. Experimental studies suggest that phonemes are encoded in the (left) primary auditory cortex (pAC) through dynamic neural activity patterns that are sensitive to specific acoustic features \cite{mesgarani2014phonetic}. 
We thus turn our attention to computational models that strive to link cortical dynamics and neural mechanisms to speech processing and, more specifically, to the parsing of speech into phonemes, syllables, and words.

\subsection{Neural models of phoneme parsing, identification, and temporal integration}

\begin{figure}[h]
\begin{minipage}[h]{1\linewidth}
\center{\includegraphics[width=0.8\linewidth]{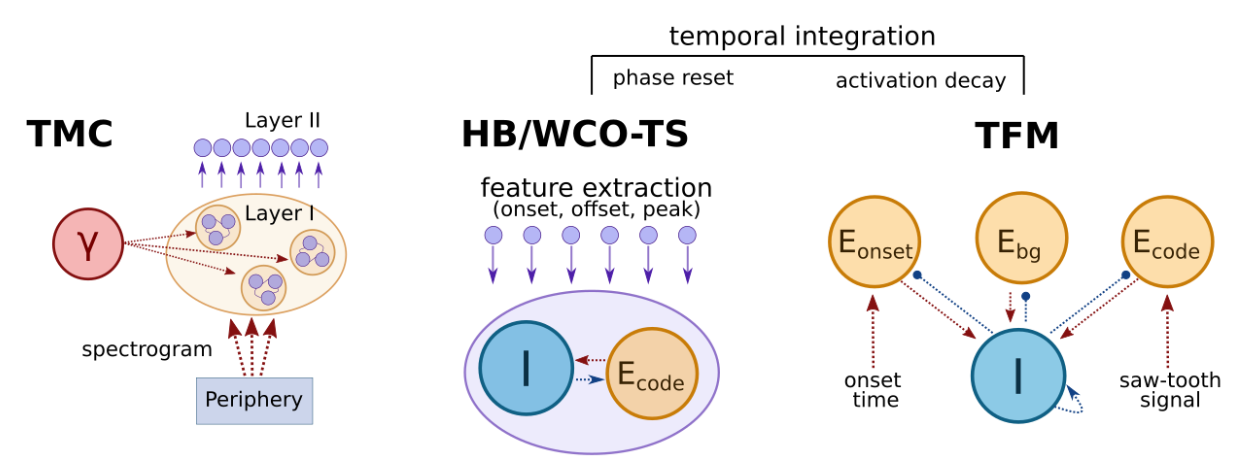}}
\end{minipage}
\caption{\textbf{Gamma-coding of phonemes is described through three mechanisms implemented as computational models.} These models vary in their methods of integrating information, which is crucial for speech processing. The TMC model relies on activity patch matching. It has a central subthreshold gamma-pacemaker (the red $\gamma$-circle) that provides the temporal modulation for the layer I neuronal voltage. Neurons within a patch receive temporally structured stimulus from a pre-assigned common frequency channel. Judiciously defining a pattern of thresholds for the neurons ensures that a patch fires in response to a specific temporal template. Layer II neurons are coincidence detectors for firing across multiple frequency channels. The phoneme is identified by spikes of neurons in layer II. So the TMC model creates a labeled-line code for the phonemes. The HB and WCO-TS models employ decay-time tuning of neural activation to label word identities as temporal activity patterns. The model structure consists of two layers: the first layer is adjusted to keep the activation and provide the dynamical base for pattern formations in the second layer, where excitatory and inhibitory populations are implemented. The TFM model network consists of three spiking excitatory populations, the onset, the code, and the background population ($E_{onset}$, $E_{code}$, $E_{\gamma}$), and one inhibitory population (I). The gamma rhythm is generated by $E_{\gamma}$-I interaction in the background population interacting with the I population. $E_{onset}$-I interactions implement phase-reset of the rhythmic activity in the whole network. $E_{code}$ receives input dependent on the phoneme shape and generates patterns of activity reflecting different phoneme identities. More details are given in the main text.}
\label{fig:fig2}
\end{figure}

To better understand these processes, various computational models have been developed to link the dynamics of neural oscillations to the parsing of speech into phonemes, syllables, and words. These models generally fall into two broad categories: those that utilize neural oscillations and dynamics to create a labeled-line code, where each phonemic unit is represented by a distinct wave of neural activity, and those that represent phonemes through specific patterns of neural activity that can be decoded from the overall neuronal response. In this section, we review several rhythm-based models that aim to explain how neural oscillations contribute to the representation and integration of phonemic information (Fig.~\ref{fig:fig2}). 

Several models are designed for word recognition based on gamma rhythmic activity. The Template-Matching Circuit (TMC) model \cite{ghitza2007towards} demonstrates how neural encoding of phonemes can be achieved through a labeled-line code using a gamma-pacemaker mechanism. The Hopfield-Brody (HB) model \cite{hopfield2000moment, hopfield2001moment} focuses on temporal sequence encoding to recognize complex spatiotemporal structures, such as spoken words. The Weakly Coupled Oscillator for Transient Synchronization (WCO-TS) model \cite{zavaglia2012dynamical} explores the role of transient synchronization among neuronal oscillators in speech recognition. The Time-Frequency Match (TFM) model \cite{shamir2009representation} utilizes gamma rhythms and stimulus-onset signals to encode phonemic information through patterns of spiking activity.

Despite their intended purpose, these models have only been tested on relatively simple datasets, including spoken digits (TMC \cite{ghitza2007towards}, HB \cite{hopfield2000moment}, WCO-TS \cite{zavaglia2012dynamical}) and saw-tooth synthetic word-like signals (TFM \cite{shamir2009representation}). These types of signals represent a limited subset of the broader range of acoustic representations found in natural speech. Given that phoneme coding has been observed to occur across low- and high-gamma rhythms \cite{mesgarani2014phonetic}, these models are more accurately described as phonetic processing models rather than true word recognition models. The limited scope of their test material does not fully encompass the complexity of word recognition, which involves more varied and intricate acoustic patterns. Moreover, effective word recognition also requires consideration of top-down contextual information flow, which helps in interpreting and predicting spoken language within its context \cite{davis2007hearing, gagnepain2012temporal, sohoglu2012predictive}. Thus, we refer to these models as phonetic processing models rather than word processing models. The following sections discuss the specific mechanisms each model proposes, the experimental evidence supporting them, and their potential applications and limitations in modeling human speech processing. 

\subsubsection{The Template-Matching Circuit (TMC) model}

One of the earliest models demonstrating neural encoding of phonemes is the Template-Matching Circuit (TMC) model \cite{ghitza2007towards}. TMC is a neuromorphic model designed to simulate phoneme recognition in speech signals by leveraging the brain’s rhythmic activity, particularly in the broad gamma frequency range (~30 Hz). The model aims to explain how neural circuits in the auditory cortex might encode phonemic information by creating a labeled-line code, where distinct phonemes are represented by specific patterns of neural activity. 

\begin{figure}[h]
\begin{minipage}[h]{1\linewidth}
\center{\includegraphics[width=0.9\linewidth]{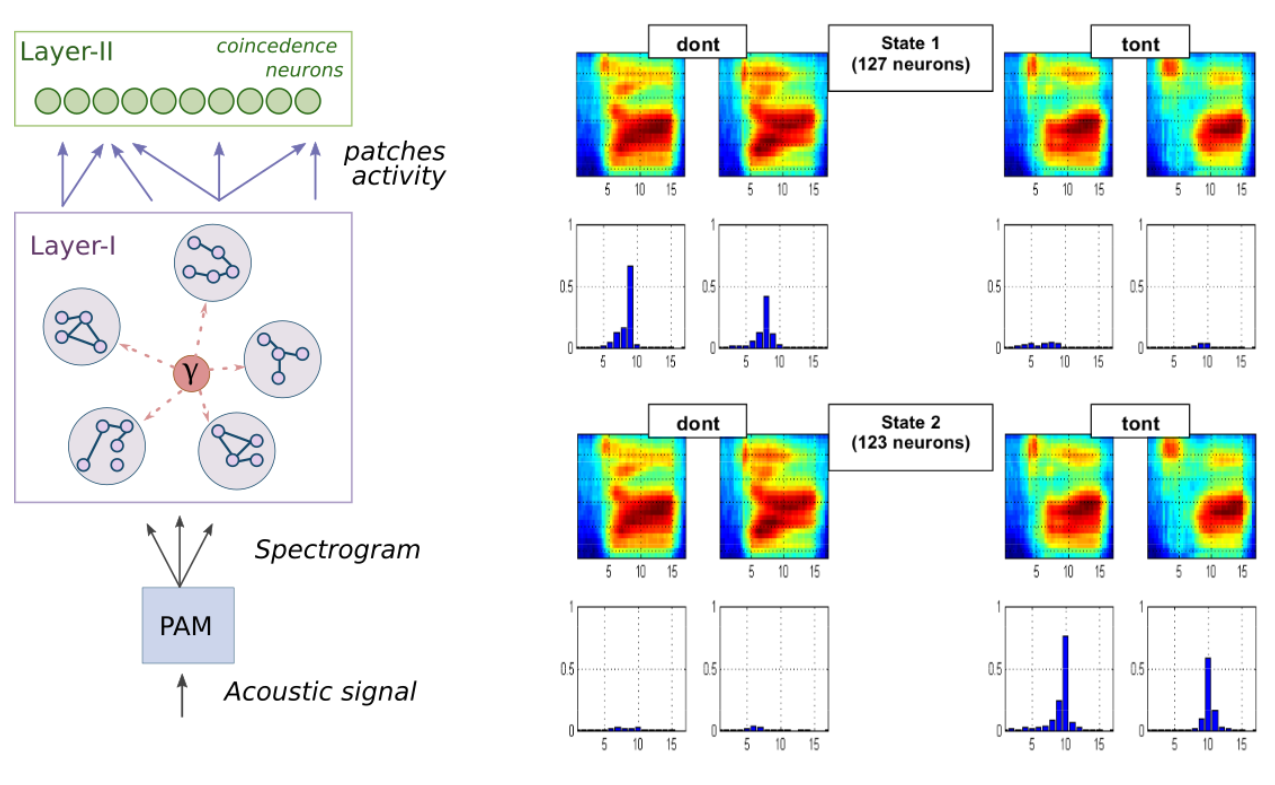}}
\end{minipage}
\caption{\textbf{Scheme of the Template-Matching Circuit (TMC) model.} The acoustic signal is transformed into a spectrogram by the PAM (peripheral auditory model) module \cite{ghitza1986auditory}. Layer-I consists of integrate-and-fire neurons with heterogeneous thresholds that are combined into patches and modulated by one gamma oscillator (red circle). These patches have acoustic features (frequencies) as an input and provide neuronal activity that is passed into Layer II, where neurons encode phoneme representations. Each Layer II neuron is "calibrated" to a specific time-frequency pattern, defined by the time evolution of frequency channels. Neurons in Layer II are also called coincidence neurons and are activated by a conjunction of specific patches in Layer-I. Right: Example of the TMC model performance in a consonant discrimination task. State 1 and State 2 represent the activity of neurons in Layer II that are most sensitive to the stimulus for two separate words: “dont” (State I) and “tont” (State 2). Two separate utterances of each word are shown. The spectrograms depict the input to the model from the PAM module neurons during the first 200 ms after stimulus onset. Below the spectrograms are time-histograms showing the number of state neurons responding to the corresponding stimulus. Adapted from \cite{ghitza2007towards}.}
\label{fig:fig3}
\end{figure}

The TMC model consists of three interconnected layers of neurons that work together to transform acoustic signals into recognizable phoneme representations (Fig.~\ref{fig:fig3}).

Layer I is composed of 2600 integrate-and-fire neurons organized into "patches." Each patch is a group of 100 neurons receiving input from a specific frequency channel of the speech spectrogram, which is generated by a Peripheral Auditory Model (PAM) that mimics signal processing in the human cochlea \cite{ghitza2007towards}. Neurons within each patch exhibit diverse temporal threshold levels and hence fire at different timepoints of the input signal. The neurons of Layer I are modulated by a common gamma oscillator, creating rhythmic patterns of activity that align with the acoustic features of the input. Layer II contains "coincidence detector" neurons that are activated by a fixed number of randomly chosen patches. The connection strengths are chosen so that the target neuron receives input from only the most active neurons in each patch. Thus, each Layer II neuron is tuned to recognize a specific temporal pattern of activation across several patches. These neurons are only activated when they receive synchronous input from the corresponding patches, effectively creating a neural code for phonemes. Hence the code for the phonemes in this model can be thought of as a labeled-line code.

The TMC model uses gamma oscillations as a timing mechanism to enhance the temporal precision of neural responses to speech signals. The synchronization provided by gamma rhythms ensures that the neurons in Layer I are selectively responsive to specific time-frequency patterns, which are then detected by coincidence neurons in Layer II. This allows the model to represent each phoneme uniquely based on the coordinated firing patterns of neurons, providing a robust coding scheme for phonemic information. A key feature of the TMC model is its resilience to noisy conditions. The gamma rhythm-based synchronization helps to filter out irrelevant background noise and maintain the integrity of phoneme recognition. In simulations, the TMC model demonstrated successful recognition of consonants in speech degraded by noise (Fig.~\ref{fig:fig3}).

The main limitation of the TMC model is that reliance on a fixed gamma frequency may not capture the full adaptability of the brain to different speech rates and varying contexts. Additionally, the model's approach to phoneme recognition does not account for the integration of diphones (combinations of two phonemes) across multiple gamma cycles, which is necessary for complex speech processing tasks. Future extensions of the TMC model could incorporate mechanisms for dynamic adaptation of the gamma frequency or additional layers of temporal integration to better handle the complexities of natural speech. For example, models like the Time-Frequency Match (TFM) have extended the TMC framework by introducing mechanisms that allow for more flexible encoding of diphone information.

\subsubsection{The Hopfield-Brody (HB) model for temporal sequence encoding}

The Hopfield-Brody model \cite{hopfield2000moment, hopfield2001moment, hopfield2004encoding}, is designed to recognize complex spatiotemporal patterns, such as spoken monosyllables, by leveraging transient synchrony among neurons. This model addresses how the brain might integrate sensory information over short intervals of time (approximately 0.5 seconds) to form a coherent perception of a "moment." The model is particularly focused on understanding how transient synchrony, or brief periods of synchronized neural firing, could serve as a mechanism for recognizing temporal sequences of sensory input. 

\begin{figure}[h!]
\begin{minipage}[h]{1\linewidth}
\center{\includegraphics[width=0.8\linewidth]{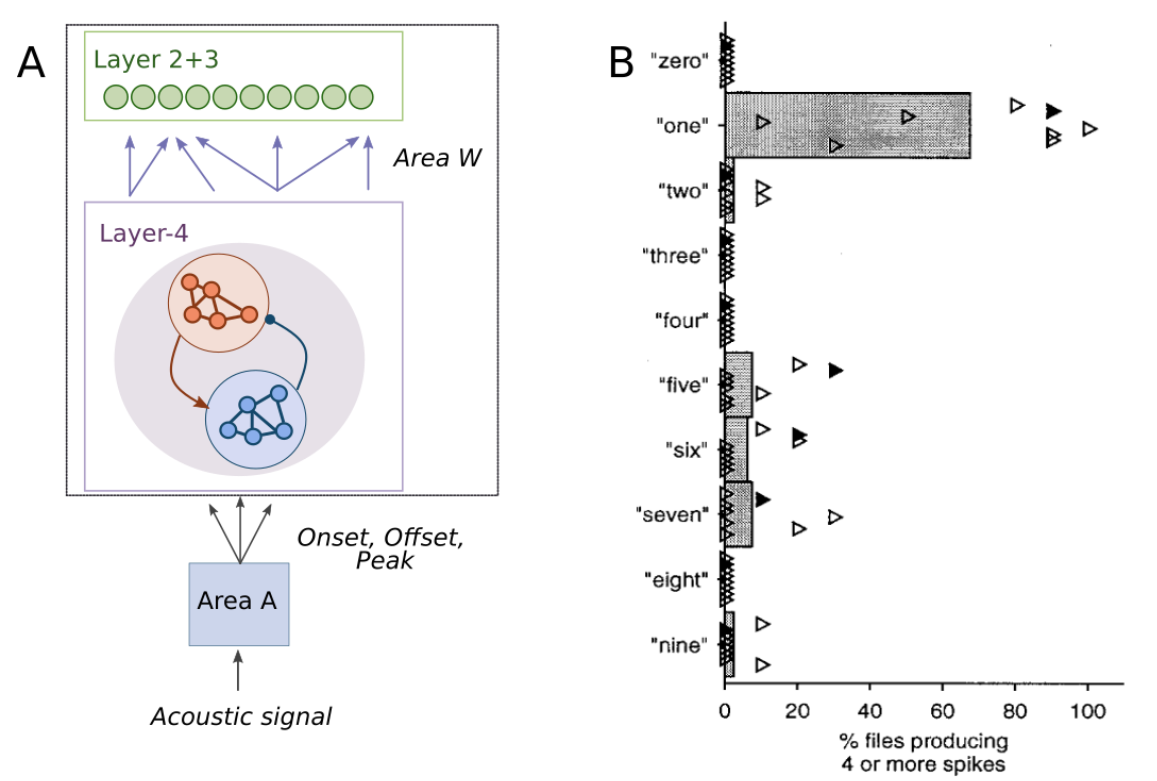}}
\center{\includegraphics[width=0.8\linewidth]{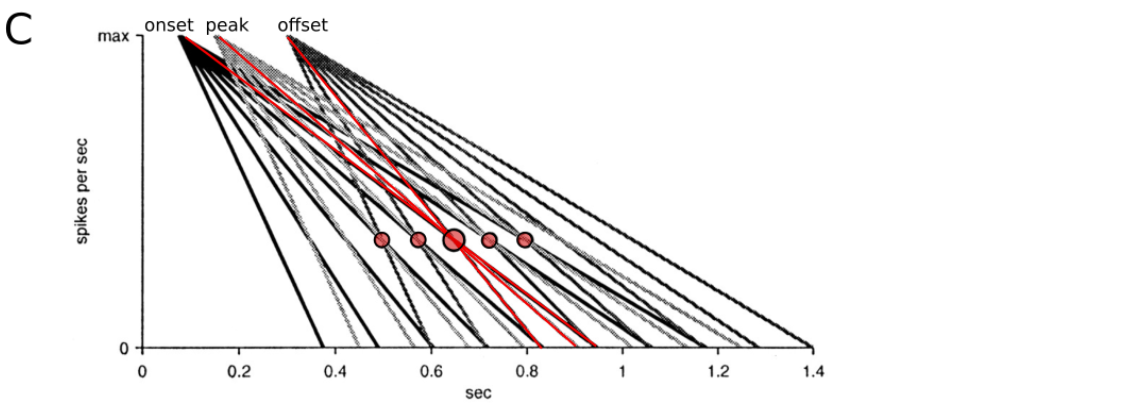}}
\end{minipage}
\caption{\textbf{HB model for temporal sequence encoding for speech recognition.} (A) Schematic illustration of the HB model. Area A is a sensory perception area that has a tonotopic structure. Projections from Area A to Layer 4 of Area W are split, preserving the tonotopy. Nominal layer $4$ includes two types of cells: excitatory (orange) and inhibitory (blue) neurons. These cells have recurrent connections and connections to cells (green) of Layer 2+3. Area A cells are frequency-tuned and respond to temporal changes in acoustic signals with a series of action potentials with a slowly decaying firing rate. Area W encodes words. For each frequency, neural subnetworks arise, which first respond with the corresponding maximum frequency and then decay with some tau constant trained on onsets and offsets. Network learning means adding new synaptic connections between excitatory and inhibitory cells necessary to activate the appropriate pattern. (B) Summary of a single gamma-neuron's responses to 10 spoken digits from the TI46 database. A trial was classified as "responding" if the gamma-neuron fired four or more spikes. Triangles represent the averages of different utterances by individual speakers. Gray bars depict the data averaged over all utterances from all speakers. The filled symbol indicates the speaker from whom the single training utterance was taken. (C) Decaying currents for three different channels: onset, offset, and peak. Red dots identify the convergence of the currents. The convergence of these three feature currents (red lines) forms an activation pattern at Layer 4, which is projected to Layer 2+3. Adapted from \cite{hopfield2000moment}.}
\label{fig:fig4}
\end{figure}

The HB model consists of a neural network with two main processing areas: Area A and Area W with Layer 4 and Layer 2+3. Area A serves as the initial sensory processing area. It includes neurons tuned to specific frequencies and types of transient acoustic features (onsets, offsets, and peaks of power in the input signal). Neurons in Area A are organized into "input channels," each corresponding to a different frequency band and type of acoustic feature detector. Each neuron in Area A produces a decaying current in response to an input event, with a range of decay rates. These currents represent the time elapsed since the input event occurred. This encoding allows the network to maintain a temporal representation of sensory inputs over brief intervals (up to 0.5 seconds).

Layer 4 of Area W receives direct excitatory input from Area A. Neurons in this layer are divided into excitatory and inhibitory, maintaining a balance between excitation and inhibition and allowing a transient synchrony. The output neurons, referred to as gamma-cells, are located in Layer 2+3. In Area W, neurons with similar firing rates due to their decaying inputs tend to transiently synchronize through synaptic coupling. This is a key mechanism for recognizing patterns: when the input pattern matches a stored template, synaptic connections enhance neuronal synchronization, leading to a collective recognition signal. Synchronization occurs when neurons receive nearly identical input patterns at the same time. Their synchronous firing is detected by gamma-cells, which signal that the target pattern has been recognized. Each gamma-cell is selective for a specific spatiotemporal pattern, such as monosyllabic/bisyllabic words. 

The HB model has been validated using simulations of speech recognition tasks. For example, in recognizing the monosyllable "one," the network demonstrates high selectivity for this word across various speakers, speeds, and noise levels. The gamma cells reliably fire when the input matches the stored template, demonstrating the model’s ability to handle the natural variability in auditory signals, aligning with observed results in behavioral experiments \cite{garvey1953intelligibility, ghitza2014behavioral}. 

The model demonstrates robustness to natural variations in the input, such as changes in speaker, speaking rate, and background noise. It achieves this through its ability to recognize patterns based on the degree of synchrony among neurons rather than on the precise timing of individual spikes. The architecture also provides invariance to time-warping and intensity changes of the input events. This is accomplished by detecting the convergence of decaying currents from different channels; when an input pattern is time-warped, the neurons' currents still converge at a different level, preserving the recognition capability.

While the HB model provides a robust framework for recognizing short spatiotemporal patterns, it also has limitations. It requires carefully tuned synaptic weights and precise transient synchrony, which may not fully capture the variability of real neural circuits. Moreover, the model focuses on short temporal sequences and may need further development to handle more complex and extended speech inputs.

Also, the model relies on explicit "onset," "peak,” and “offset” feature signals that set the temporal windows for pattern recognition. By using these three features, the model essentially reduces the phonetic pattern representation to sawtooths (as in \cite{shamir2009representation}). This thus works for simple acoustic patterns but is clearly not sufficient for complex words and large word dictionaries. Presumably, relaxing the need for explicit onset and offset signals could be easily addressed by including slow rhythms that are described below.

Finally, the HB model has a strict assumption that synchronization requires an exquisitely detailed balance of excitation and inhibition in the spiking network. However, this is not the sole mechanism enabling network synchronization. To overcome the requirement for strict excitation-inhibition balance and make the model more generalized, a more abstract model of phase synchronization has been proposed in the work of \cite{zavaglia2012dynamical}.

\subsubsection{Weakly Coupled Oscillator Model of Transient Synchronization (WCO-TS)}

The Weakly Coupled Oscillator for Transient Synchronization (WCO-TS) model \cite{zavaglia2012dynamical} offers a framework for understanding gamma activity in the auditory cortex during speech recognition. This model extends the HB-model \cite{hopfield2000moment, hopfield2001moment} by focusing on transient synchronization of neural oscillators to detect acoustic patterns in speech. It proposes that recognition events are marked by bursts of gamma activity, which align with observed ECoG data \cite{canolty2007spatiotemporal} during auditory word recognition.

The processing starts with a filtering of auditory inputs into several frequency bands. Then, similarly to the HB model, detectors identify three main types of features: onsets, offsets, and peaks of power across these bands. The WCO-TS model consists of a network of phase oscillators that are weakly coupled \cite{hoppensteadt1997weakly}, allowing their phases to align when they receive similar inputs. Synchronous firing in the network elicits gamma activity, measured as LFPs. The time of the strongest synchronization leads to an increase in the LFP amplitude and gamma power. This synchronization across oscillators indicates a "recognition event" when a specific spatiotemporal pattern matches a stored template. 

The WCO-TS model has invariance to time compression and stretching, meaning it can recognize patterns despite changes in the speed at which words are spoken. This is achieved by manually adjusting the oscillation frequencies to match the pattern of incoming stimuli.

\begin{figure}[h]
\begin{minipage}[h]{1\linewidth}
\center{\includegraphics[width=0.9\linewidth]{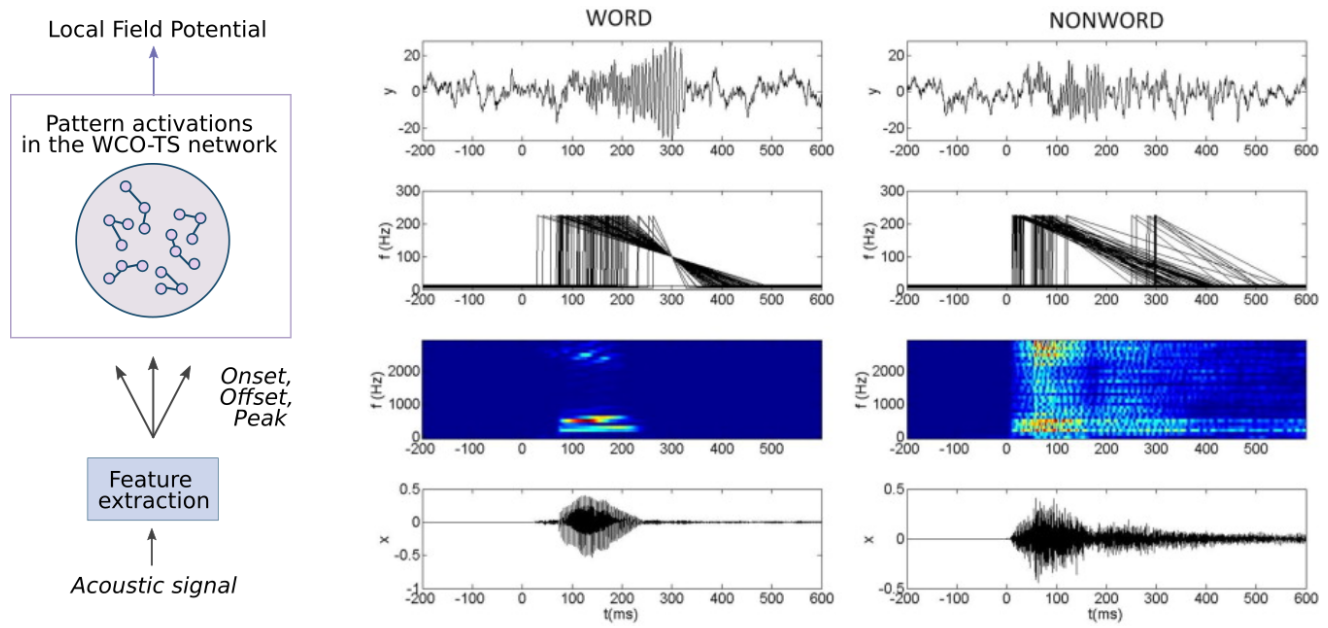}}
\end{minipage}
\caption{\textbf{Schematic of the model Weakly Coupled Oscillator of Transient Synchronization (WCO-TS).} Left: The WCO-TS model includes feature extraction, pattern activation in the network of weakly coupled oscillators, and gamma synchronization as in the activity of the local field potential. Right: Example of a model response to a word and nonword stimuli, with gamma bursts indicating recognition events. The top panel shows the LFP signal (y); the second panel shows detection of feature-occurrence time; the third panel is for auditory spectrograms; and the bottom panel x(t) is auditory input. Adapted from \cite{zavaglia2012dynamical}.
}
\label{fig:fig5}
\end{figure}

The model is fitted to ECoG data, demonstrating its ability to predict high gamma activity observed in response to word versus non-word stimuli. Of note, the dynamical model was used only as a forward model for synthetic spectrograms to be compared with ECOG data recorded in human subjects who listened for spoken digits. Figure~\ref{fig:fig5} demonstrates the main steps of the WCO-TS model.

The WCO-synchronization model extends the original HB model \cite{hopfield2000moment, hopfield2001moment} by applying the synchronization mechanism in a network of weakly coupled oscillators (WCO). In the HB model, transient synchronization is achieved through E-I balance, which is not a confirmed mechanism. And the WCO-TS model aims to generalize the model to a more neutral phenomenological level. Thus, this feature allows for future studies of the mechanisms underlying feed-forward speech perception and temporal sequence processing through synchronized neural rhythms. While not showing that specific words are detected in the WCO model explicitly, the authors do show that a feature-time-occurrence code can be used to identify words algorithmically by a first nearest neighbor classifier \cite{duda1973pattern}. While the authors appeal to synaptic plasticity as a potential mechanism for structuring the WCO network clusters, they do not explicitly implement this learning; rather, the network is engineered by hand for a small number of predetermined input stimuli. 

\subsubsection{The Time-Frequency Match (TFM) model for gamma-based sequence encoding}

The TFM model \cite{shamir2009representation} presents a computational framework for understanding how the auditory cortex might encode time-varying stimuli, such as speech, using neural oscillations in the gamma frequency range. The model suggests that gamma oscillations play a direct role in representing stimuli whose duration exceeds a single gamma cycle. It is designed to explain how neural circuits can achieve robust speech perception, even under variations in speech timing or noise.

\begin{figure}[h]
\begin{minipage}[h]{1\linewidth}
\center{\includegraphics[width=0.9\linewidth]{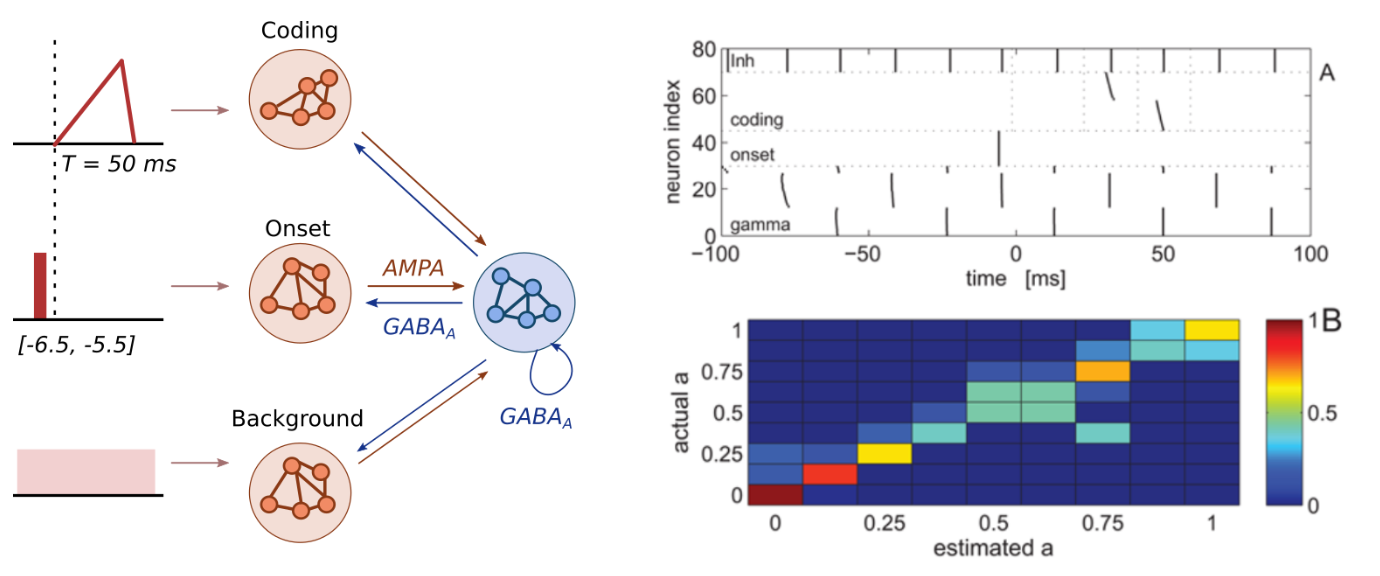}}
\end{minipage}
\caption{\textbf{Schematic of the Time-Frequency Match (TFM) model.} The model consists of excitatory (orange) and inhibitory (blue) subpopulations. The PING mechanism generates gamma rhythm activity. The excitatory populations are divided into onset (synchronized with stimulus onset), coding, and background populations. The coding population encodes time-dependent sawtooth input. Right: (A) Example of the TFM model activity during stimulus processing. The onset signal occurs at t = 26.5 ms. Neuron indexes 1-30, 31-45, 46-70, and 71-80 correspond to the background, onset, coding, and inhibitory subpopulations, respectively. (B) Confusion matrices for discriminating nine alternative shapes of sawtooth signals. Adapted from \cite{shamir2009representation}.}
\label{fig:fig6}
\end{figure}

The TFM model features a pyramidal-interneuron gamma (PING) network \cite{whittington2000inhibition, buzsaki2012mechanisms, borgers2003synchronization, dumont2019macroscopic}, where interconnected populations of pyramidal neurons and interneurons generate activity in gamma rhythm. The TFM utilizes speech envelopes and gamma rhythm to encode phonemes that occur in the speech stimuli (Fig.~\ref{fig:fig6}) into patterns of spiking activity.

In the TFM model, the network comprises 70 pyramidal neurons and 10 interneurons. The interneuron populations are common, and the pyramidal neurons are split into three populations that play the following key roles:

\begin{enumerate}
    \item \textit{Background subpopulation} generates a self-sustaining gamma rhythm through the PING mechanism, creating background activity in the network. 
    \item \textit{Onset subpopulation} resets the gamma oscillation phase by an artificially generated stimulus onset. The onset signal allows the network to align its activity with speech stimuli for continuous gamma coding of the signal. In the model, the onset time occurs every three gamma cycles.
    \item \textit{Coding Subpopulation} consists of 25 pyramidal neurons, and each neuron receives a sawtooth input current. All the feed-forward weights are uniform, but the background constant currents to these neurons are designed so that each form of input signal (i.e., each sawtooth shape) evokes the response of a defined subgroup of neurons in the coding subpopulation. The signal is characterized by different rates of rise-to-peak (e.g., the envelope). Neurons that are tuned to react with different sensitivities are activated at different rates, producing responses and signaling relative peak timing. Thus, the discretization of the phoneme representations is carried out due to the variability in the sensitivity of neurons as defined by the level of the background constant current.
\end{enumerate}

The TFM model incorporates not only the gamma rhythm, as in the TMC model, but also the stimulus onset signals for more robust ongoing encoding by signal discretization. Phonemes in the TFM model are associated with patterns of gamma rhythm activity that represent the response of the network (coding subpopulation). Since neurons in the coding population are sensitive to different input signals, phase resetting at onset times provides the ability to fire at the correct timings. Thus, the onset subpopulation helps to encode phonemic boundaries during online recognition of a pseudo-rhythmic acoustic signal.

The model read-out is done by an artificial template matching algorithm. Thus, the model does not implement a neuronal readout mechanism, and the authors propose that neuronal integrators could be used to accumulate information over the period from one onset to the next. Such models can be based on calcium current in the dendrites \cite{loewenstein2003temporal}, long-term potentiation and depression \cite{buonomano2000decoding}, or slow voltage-dependent conductances \cite{hooper2002computational}.

Numerical tests demonstrate that the model performance can be resilient to input rate variations and neural noise, presumably because the onset pulses are able to reset the activity and effectively set the temporal boundaries for processing the phonetic identity of the signal. Numerical experiments also show that the onset signals are necessary for effective speech encoding, as without them the model is unable to parse phonemes. However, the TFM model has some limitations and areas for future development. Specifically, it uses a saw tooth input current instead of speech signals for testing, and the mechanism for obtaining stimulus onset signals remains underexplored. Also, in the presented model only one frequency channel is considered, yet in speech, multiple frequency channels carry relevant sound information envelopes. Taking this into account would require both an increase in the number of neurons in the coding subpopulation and a change in the code structure. One would have to decide whether the coding subpopulation is restricted to a single frequency channel or integrates all channels with similar sawtooth shapes. It is interesting to point out that in the TFM, the decision to incorporate synchronization was inspired by the model of Hopfield and Brody \cite{hopfield2000moment, hopfield2001moment}.

\subsubsection{The Tripod network for phoneme sequence}

In the models we have discussed so far in this section, the central role is attributed to the variability of thresholds (TMC and TFM) or decay times (HB and WCO-TS), which facilitate the integration of temporal signals. Notably, the calibration of these thresholds was performed in these models manually. In the study \cite{quaresima2023dendrites}, the focus is on the role of dendrites and synaptic plasticity in learning to identify words based on temporally ordered stimulus sequences. The findings demonstrated how a spiking network of dendritic neurons could learn stable associations between phonemes and words, enabling the extraction of correct word sequences from phonemic stimuli. To achieve this, the Tripod network model \cite{quaresima2023dendrites} employs a dendritic memory mechanism that sustains depolarized states, linking sequences of incoming phonemes.

\begin{figure}[h]
\begin{minipage}[h]{1\linewidth}
\center{\includegraphics[width=0.9\linewidth]{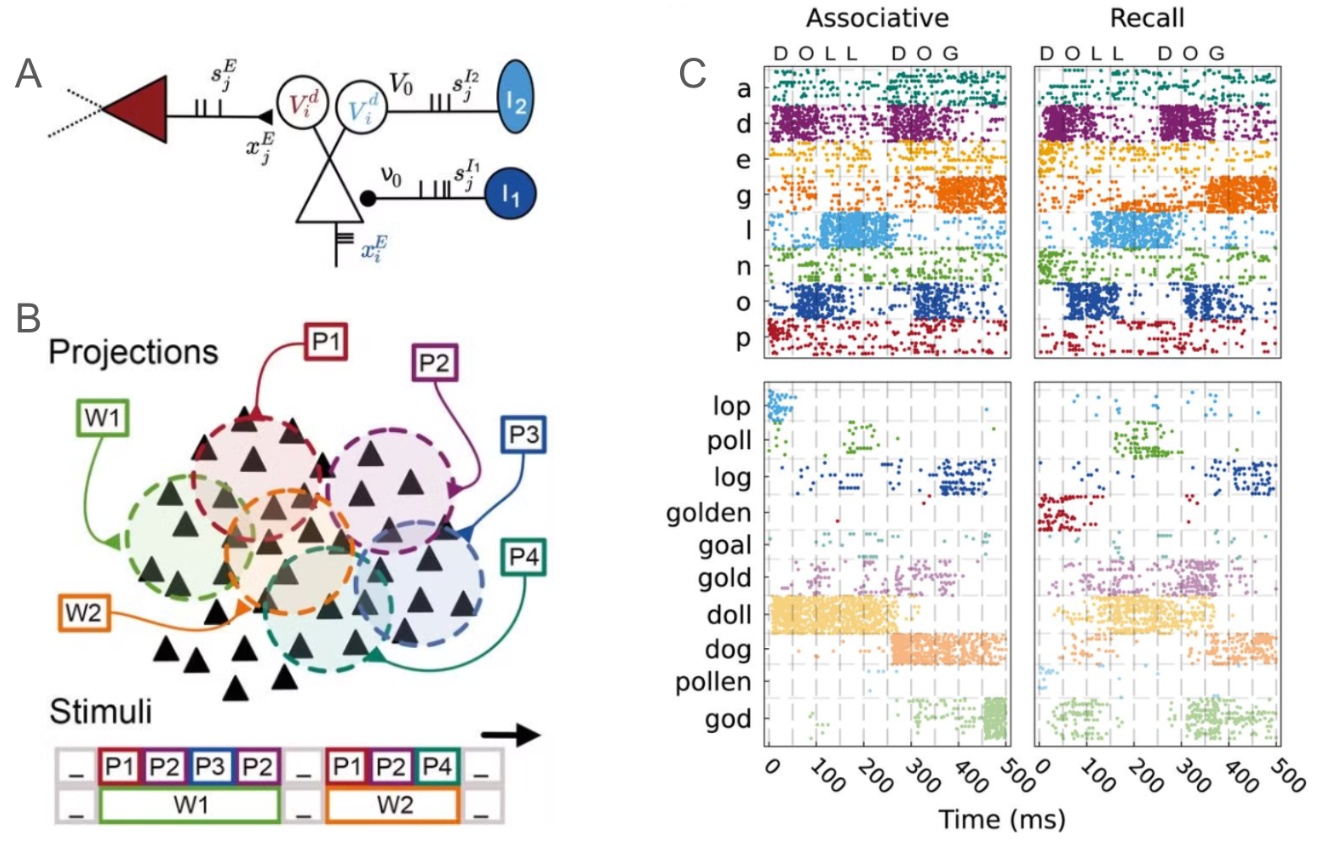}}
\end{minipage}
\caption{\textbf{The Tripod network.} (A) The Tripod neuron model. It has 3 compartments: 2 dendrites (circles) and soma (triangle). A voltage-based STDP in the glutamatergic synapses between excitatory neurons (Vd is a membrane potential of dendritic compartments). The inhibitory synapses (sI1 and sI2) onto the tripods are subject to iSTDP. A pre-synaptic activity (xE) drives both excitatory and inhibitory plasticity. (B) The network consists of 2000 tripod neurons, 175 fast and 325 slow-spiking inhibitory neurons. Each neuron connects randomly to 20$\%$ of the network via dendritic synapses through vSTDP. The network maintains stable, continuous activity through iSTDP. During training, the network is divided into overlapping activation regions corresponding to words (W1, W2) and phonemes (P1-P4). (C) Rasterplots of network activity during the associative (learning) and recall phases. In the associative phase, the network is exposed to sequences of words, with both word projections and corresponding phoneme sequences being activated simultaneously. This co-activation results in the development of overlapping engrams among the coactive neurons, establishing associations between phonemes and words. During the recall phase, the network is presented with phoneme sequences alone, which prompts the reactivation of the word populations based on the associations formed during learning. Adapted from \cite{quaresima2023dendrites}.}
\label{fig:fig7}
\end{figure}

Each tripod-neuron within the network comprises three compartments: two dendrites and one soma (Fig.~\ref{fig:fig7}A). The coordinated dynamics of NMDA receptors and voltage-gated channels on the dendrites establish a dendritic memory: a depolarized dendritic voltage plateauthat lasts approximately 100 milliseconds. The model includes 2,000 Tripod excitatory neurons and 175 fast and 325 slow inhibitory single-compartment interneurons (Fig.~\ref{fig:fig7}B). Each neuron in the network independently connects to 20$\%$ of other neurons via conductance-based synapses, where voltage-dependent spike-timing-dependent plasticity (vSTDP) occurs. Importantly, the excitatory synapses and the slow inhibitory ones target the dendrites on the tripods, and the fast inhibition targets the somata. Excitatory plasticity allows the formation of engrams within the cellular assemblies.

Noise in the model drives the neurons to exhibit spontaneous activity. The continuously active regime is stabilized by inhibitory spike-timing-dependent plasticity (iSTDP) between inhibitory neurons and the somas of excitatory cells. Additionally, voltage-dependent iSTDP is maintained between slow-firing inhibitory neurons and targeted dendritic compartments, allowing for more precise control of dendritic nonlinearities and facilitating computations on the dendrites.

The model operates in two phases: associative and recall (Fig.~\ref{fig:fig7}C). During the associative phase, the network learns sequences of words separated by brief intervals of silence as an input. For each word, the network receives strong stimulation of word projections along with the corresponding phonemes in sequence. The simultaneous activation of words and phoneme sequences leads to the formation of overlapping engrams among coactive cells through vSTDP, thereby establishing phoneme-word sequence associations during learning. In the recall mode, the network is stimulated by phoneme sequences. As a result of the learned phoneme-word associations, words are presynaptically connected to a subset of excitatory neurons, which, in turn, stimulate dendritic compartments with intense presynaptic excitation.

Thus, the constructed model suggests a biologically plausible mechanism for learning pattern sequences via dendritic plasticity. The tripod network can recognize words (phoneme sequences), where dendritic computations provide for storage and recall of words (i.e., phoneme sequences) in short-term memory of the order of 100 ms and stable long-term memory via synaptic plasticity. In fact, model comparison simulations show that without the dendrites, word recognition fails as soon as there are overlapping phonemes between the words.

Although the model is biologically inspired, its assumptions related to dendritic memory and NMDA receptor kinetics might not be applicable across different neuronal types and cortical regions. This may limit the model's applicability to actual brain circuits, as dendritic memory and plasticity rules can vary widely \cite{mel2017synaptic, poirazi2020illuminating}. Furthermore, the model's reliance on retaining synaptic and dendritic memory might cause scaling challenges. As the lexicon size and phonological overlap increase, the model's ability to encode and recall sequences with high accuracy collisions. In fact, authors already recognize some of these constraints by pointing out that only short words can be successfully recognized due to the short time scales of the dendritic plateaus. Larger datasets or complex linguistic patterns highlight the network's constraints. Moreover, the Tripod network focuses solely on phoneme-level processing without explicitly incorporating syllable-level segmentation. It does not address the hierarchical nature of speech processing, which may limit its applicability to more complex aspects of speech perception.

\subsection{Speech segmentation and syllable onset detection with oscillations}

\begin{figure}[h]
\begin{minipage}[h]{1\linewidth}
\center{\includegraphics[width=0.8\linewidth]{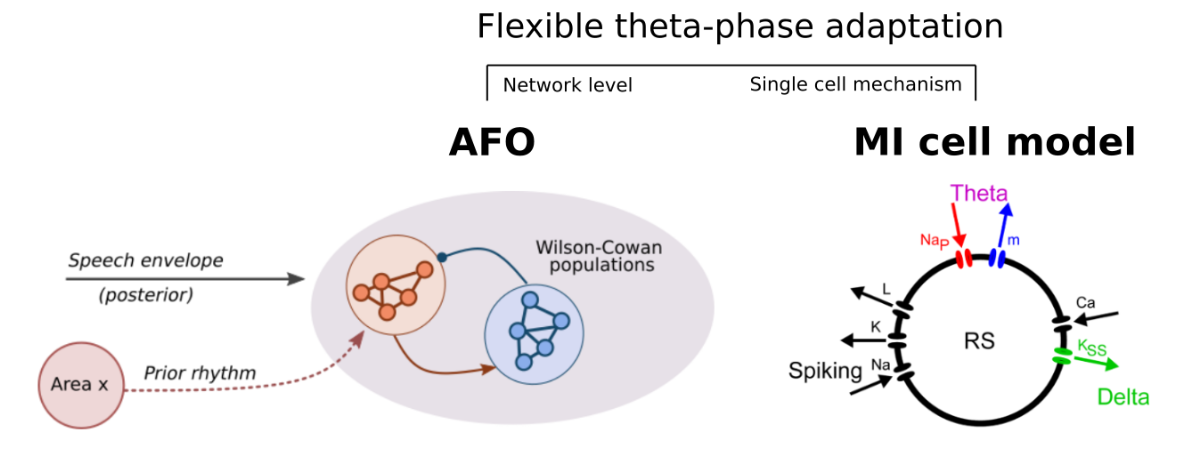}}
\end{minipage}
\caption{\textbf{Two modeling scales for flexible theta-phase adaptation.} The Adaptive Frequency Oscillator model (AFO) describes this phenomenon at the network level \cite{doelling2023adaptive}. The MI cell model delves into the single-cell mechanisms \cite{pittman2021differential}. See figures below for detailed descriptions of each model.}
\label{fig:fig8}
\end{figure}

The models depicted above suggest that synchronization plays a crucial role in phoneme encoding and phonetic sequence identification. The synchronization with speech rhythm allows for signal segmentation and information integration within segments. 

Currently, there is increasing evidence for synchronization between brain rhythms and the speech envelope, although most of it pertains to theta-rhythm synchronization with the syllabic rhythm \cite{aiken2008human, gross2013speech}. Analysis of speech in various languages (European, Asian, and others) and under different conditions (audiobook reading, listening to isolated sentences, interviews, and dialogues) has revealed a common syllable rate \cite{ding2016temporal, tilsen2013speech, varnet2017cross} in the 3-8 Hz range with a mean around 5 Hz, corresponding to the theta rhythm of the brain. The observed synchronization of the syllabic rhythm with auditory cortical theta rhythm led to the hypothesis of a functional role of the theta rhythm in speech segmentation into syllables, and more specifically, for detecting the onset of syllables in the pseudo-rhythmic speech signal \cite{giraud2012cortical}. Questions related to the origin of the speech envelope and synchronization mechanisms are subjects of active research and discussions in the fields of neurophysiology and psycholinguistics and can be partially elucidated through mathematical modeling. Particularly, which specific aspects of the synchronization between theta and gamma rhythms, namely frequency, amplitude, or phase, are most crucial in speech perception and are still debated. The following section describes those models that investigate mechanisms for theta rhythm segmentation of syllables.

\subsubsection{Adaptive frequency oscillator (AFO) model}

\begin{figure}[h]
\begin{minipage}[h]{1\linewidth}
\center{\includegraphics[width=1\linewidth]{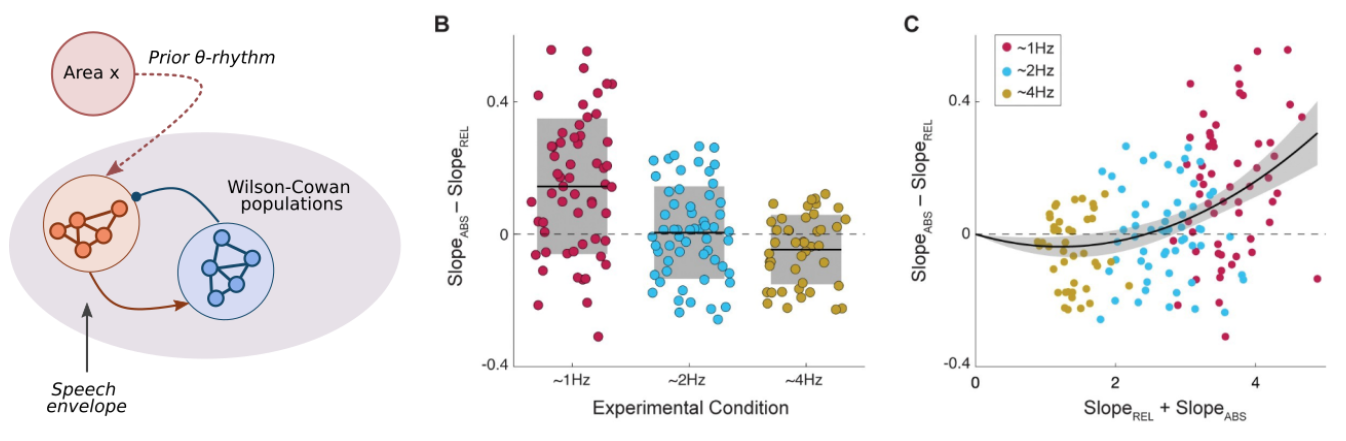}}
\end{minipage}
\caption{\textbf{The Adaptive Frequency Oscillator model (AFO).} (A) AFO scheme. Oscillations are generated by populations of interacting excitatory and inhibitory neurons. Neuronal networks are described using the Wilson-Cowan model. The excitatory population receives the signal of the speech envelope with a certain coupling gain. Area x activity controls the frequency of the oscillator to match the mean period of the envelope. (B) AFO performance is consistent with the experimental results. The discrepancy in slopes is derived from fitting logistic regressions based on answers computed using the Duration (SlopeABS) and Rhythm (SlopeREL) algorithms relative to the phase of the oscillator at the probe time. (C) The difference in slopes between the two logistic regressions plotted against the mean slope across algorithms. The black line represents a second-degree polynomial fit, with the shaded gray line indicating the 95$\%$ prediction interval. Simulated data replicates human behavior, aligning with relative timing for inferior performance and matching stimulus rate trends. Adapted from \cite{doelling2023adaptive}.}
\label{fig:fig9}
\end{figure}

The first key question is whether synchronization can serve as a mechanism for temporal phonetic prediction even in the absence of perfect isochrony in speech signals. Mechanistic models are needed to demonstrate whether oscillations can adjust to correctly phase-align despite speech warping. Several models above purported to be rate-invariant. The Adaptive Frequency Oscillator (AFO) model \cite{doelling2023adaptive} is capable of dynamically adjusting its oscillatory timescales via a background input that increases during the sound presentation, thereby increasing the oscillation frequency (Fig.~\ref{fig:fig9}). The authors proposed the AFO as a potential neural mechanism underlying the synchronization of speech rhythm and neural activity. Data shows that synchronization with a pseudo-rhythmic signal can be absolute or relative \cite{teki2011distinct}. While absolute synchronization relies on an external signal and the precise monitoring of previous intervals, relative synchronization relies on the ongoing oscillations whose endogenous frequency is modulated by extra noise. The authors \cite{doelling2023adaptive} argue that both mechanisms reflect human behavior and can be implemented by a single dynamical neural network, namely a Wilson-Cowan model \cite{wilson1972excitatory} that switches between absolute and relative synchronization. This is achieved by varying coupling coefficients in the model of the feedforward signals providing the speech envelopes. For example, different individuals are modeled as different coupling constants, from the speech envelope to the AFO network.

The AFO model bridges oscillatory and Bayesian mechanisms, where the relative synchronization algorithm is associated with a prior rhythm and absolute synchronization with an observed acoustic input \cite{doelling2023adaptive}. These mechanisms produce two estimates that are combined as a Bayesian model to infer a prediction of the next time interval. The authors argue that the absolute synchronization mechanism can be implemented through rhythmic activity.

The AFO suggests that oscillatory processes in the brain may provide prior information about the speech rhythm. While in the previous models, speech onsets needed to be explicitly signaled, in AFO, the activity that evokes the rhythm could be adjusted by, e.g., the motor areas (Area x) \cite{assaneo2021speaking}. 

\subsubsection{The onset-by-motor-prediction (OMP) model}

\begin{figure}[h]
\begin{minipage}[h]{1\linewidth}
\center{\includegraphics[width=0.8\linewidth]{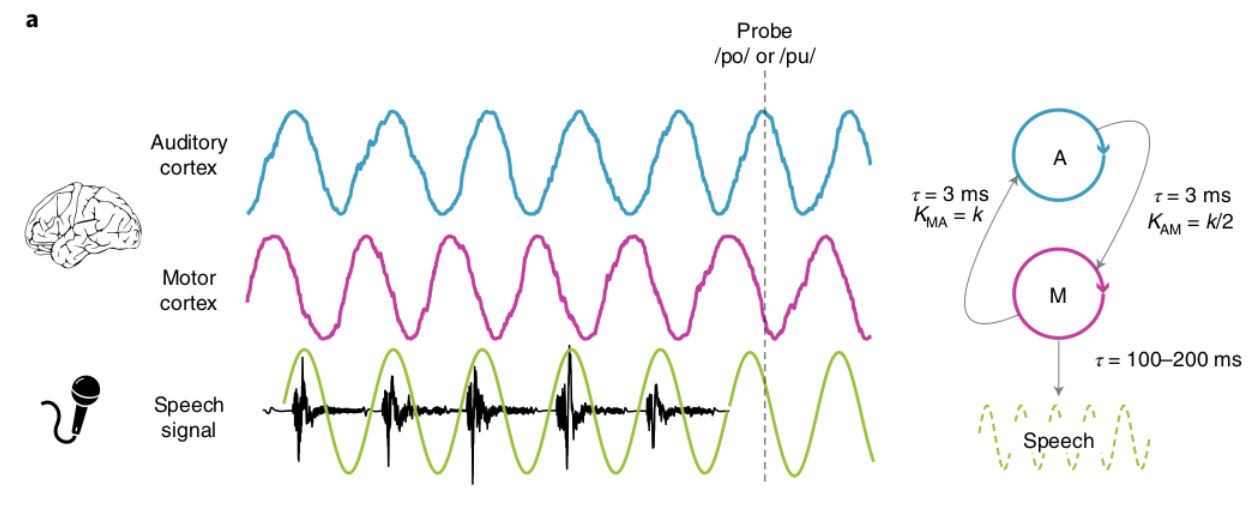}}
\center{\includegraphics[width=0.8\linewidth]{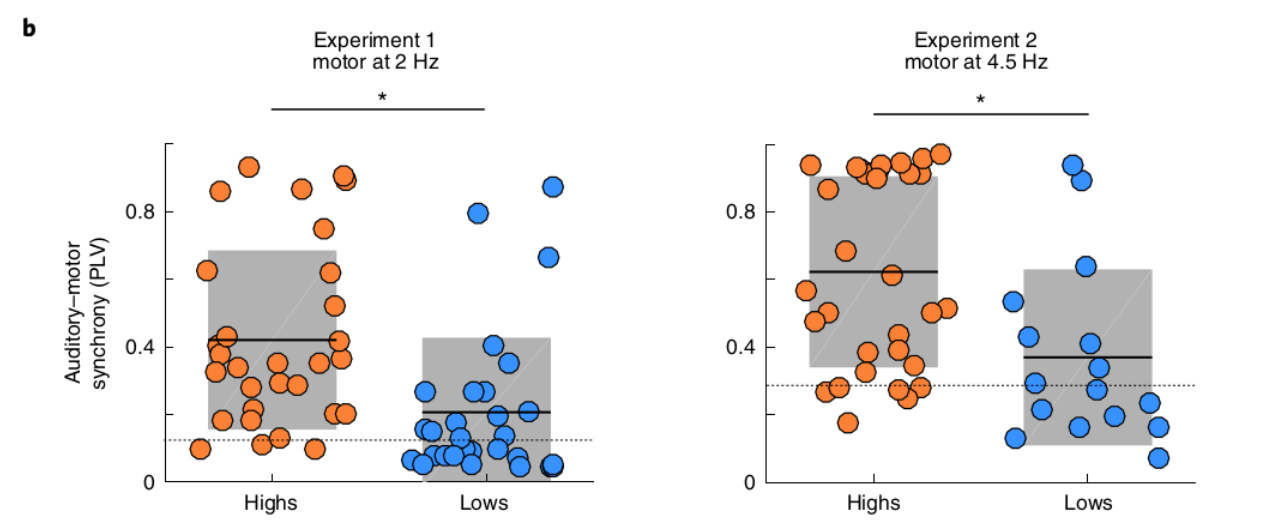}}
\end{minipage}
\caption{\textbf{The onset-by-motor-prediction model.} (A) Schematic. Motor and auditory cortical activity (blue and magenta traces, respectively) are modeled as a duo of coupled-phase oscillators. The oscillatory dynamic of the produced speech is implemented as motor cortex activity with a fixed phase lag representing the time delay observed between muscle activity and the onset of the produced speech. Motor cortex is modeled through equations representing an inhibitory-excitatory network, and the excitatory population receives the auditory cortex activity as input. (B) The auditory and motor activity synchronization obtained by simulating the different experimental conditions with the numerical model. Dots represent the participant’s degree of auditory-motor synchrony (high and low synchronizers). Adapted from \cite{assaneo2021speaking}.}
\label{fig:fig10}
\end{figure}

The OMP model elaborates on this hypothesis by proposing that temporal prediction in the context of speech segmentation arises from speech production rhythms \cite{assaneo2021speaking}. The model explores the interaction between two oscillatory populations, each identified with the auditory and motor cortices (Fig.~\ref{fig:fig10}). Each population is described using a Kuramoto phase oscillator \cite{yeung1999time}, which explicitly tracks the phase dynamics of neural activity.

The study shows how the motor cortex can predict the speech temporal characteristics and how predictions are transmitted to the auditory cortex. This model implies that the motor cortex carries out a preliminary analysis of temporal structures in speech and assists in the perception of speech segments. In order to match measurements on synchronization between the motor and the auditory cortices in individual subjects, the authors optimize the coupling between the two oscillatory populations. Consistently, behavioral data show that individual subjects can be clustered into low- and high-synchronizers. People with stronger audio-motor coupling had better speech discrimination and synchronization abilities. 

Predictive modulation of the auditory brain circuitry by the motor cortex is supported by several recent behavioral data \cite{morillon2015predictive, morillon2019prominence, keitel2018perceptually}. This model thus provides a new perspective on possible neural mechanisms involved in speech segmentation and supports temporal prediction by exploring the interaction between the auditory and motor cortices.

\subsubsection{Biophysical modeling of flexible phase synchronization: MI cell}

\begin{figure}[h]
\begin{minipage}[h]{1\linewidth}
\center{\includegraphics[width=0.8\linewidth]{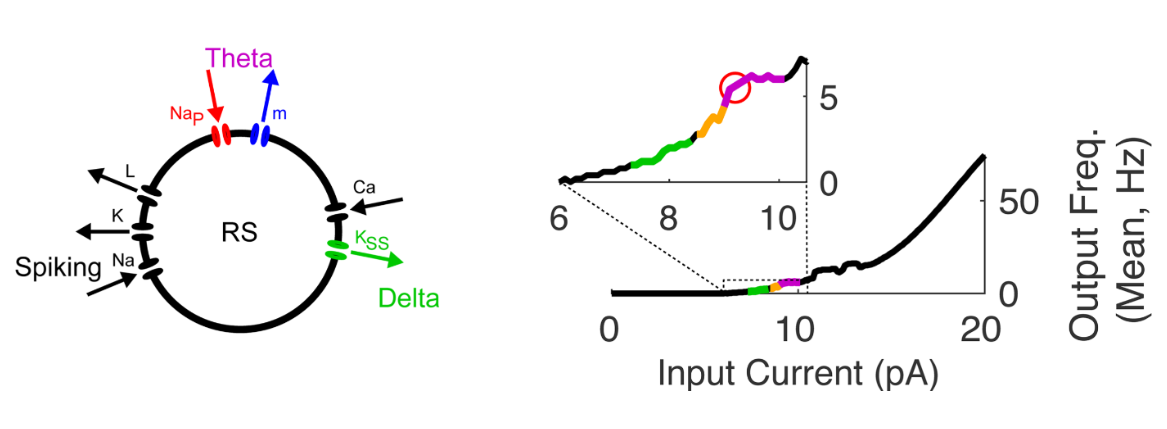}}
\end{minipage}
\caption{\textbf{Scheme of the MI cell model for flexible theta-phase adaptation.} Left: the currents color-coded according to the timescale of inhibition ($\delta$ in green, $\theta$ in purple). Right: A frequency-current curve shows the transition of spiking rhythmicity through $\delta$ and $\theta$ frequencies as $I_{app}$ increases ($\delta$ in green, $\theta$ in purple); the red circle indicates the point on the FI curve at which $I_{app}$ was fixed, to give a $7$ Hz firing rate. Arrows indicate the direction of currents (inward or outward). Adapted from \cite{pittman2021differential}.}
\label{fig:fig11}
\end{figure}

To explore the specific neurophysiological mechanisms that underlie flexible phase synchronization of neural activity with pseudorhythmic input, a smaller-scale biophysical neural-circuit model was developed by Pittman-Polletta and colleagues \cite{pittman2021differential}. This model consists of a minimal two-neuron circuit: a regular spiking pyramidal neuron and an inhibitory interneuron (Fig.~\ref{fig:fig11}). Each one is described by a detailed biophysical model that takes into account multiple ion currents. Each cell is modeled as a single compartment with Hodgkin-Huxley dynamics. First, the frequency flexibility in phase-locking can be implemented by a regular spiking (RS) cell alone, provided that in addition to the fast spiking currents, it also contains several voltage-dependent slower currents: the muscarine-sensitive non-inactivating potassium current (m-current), persistent sodium, calcium, and super-slow potassium (calcium-activated potassium) currents.

This model served to demonstrate which ions and ion currents, as well as what type of modulation, influence the dynamics of cortical theta oscillators. Flexible phase locking is possible through an interaction between slow and superslow K currents. It also depends on the subthreshold oscillations in the theta frequency range that are activated by the m-current and its interactions with the persistent sodium current. This allows neurons to flexibly adjust to the speech rhythm without exceeding their proper frequency range. Thus, this modeling sheds light on the role of individual cells with specific dynamics and ion currents in the dynamics of speech processing. The work demonstrates the feasibility of implementing mechanisms for flexible adaptation to changes in speech signals at the single-cell level. 

The authors also analyzed phase locking in the RS neuron when it is coupled with an inhibitory neuron and showed that flexible phase locking was possible in RS cells, but within a more restricted frequency range. This simulation showed that the m-current in the RS neurons and the super-slow potassium current both concurrently control the phase locking frequency range even in the presence of synaptic inhibition. 

The model relies on assumptions on the presence and dynamics of intrinsic and synaptic inhibitory currents. For example, it posits that the dynamics of cortical theta oscillations stem from single cell properties and are affected by intrinsic currents rather than synaptic ones. However, this assumption may not apply to all cortical areas or neuron types. Also, the models utilized are single-cell oscillators that may not account for network-level interactions or connectivity patterns found in cortical networks. As a result, the model is not well positioned to consider emergent network dynamics, which may impact phase-locking and speech segmentation at the population level (e.g., see \cite{dumont2019macroscopic}). Furthermore, the model parameters, membrane properties, and current dynamics are fixed and are inflexible to changes in input features or context; it would be interesting to see how the model can generalize to varied speech rates or noisy conditions.

\subsection{Theta-gamma code for speech parsing}

In the previous section, we reviewed models of neural gamma coding for phoneme representations, followed by models segmenting speech into syllables using theta rhythm. The combination of these models with theta-gamma code for processing speech stimuli in the auditory cortex seems a promising way forward \cite{giraud2012cortical}.

The proposed integration involves the coexistence and inter-coupling of the theta and gamma oscillations \cite{fontolan2013analytical, engel2013intrinsic}, where the theta rhythm segments speech into syllabic units and the gamma rhythm contributes to encoding phonemic information within those units. This combined theta-gamma mechanism offers a comprehensive approach to understanding the intricate processing of speech in the auditory cortex.

\subsubsection{The TEMPO model}

\begin{figure}[h]
\begin{minipage}[h]{1\linewidth}
\center{\includegraphics[width=0.8\linewidth]{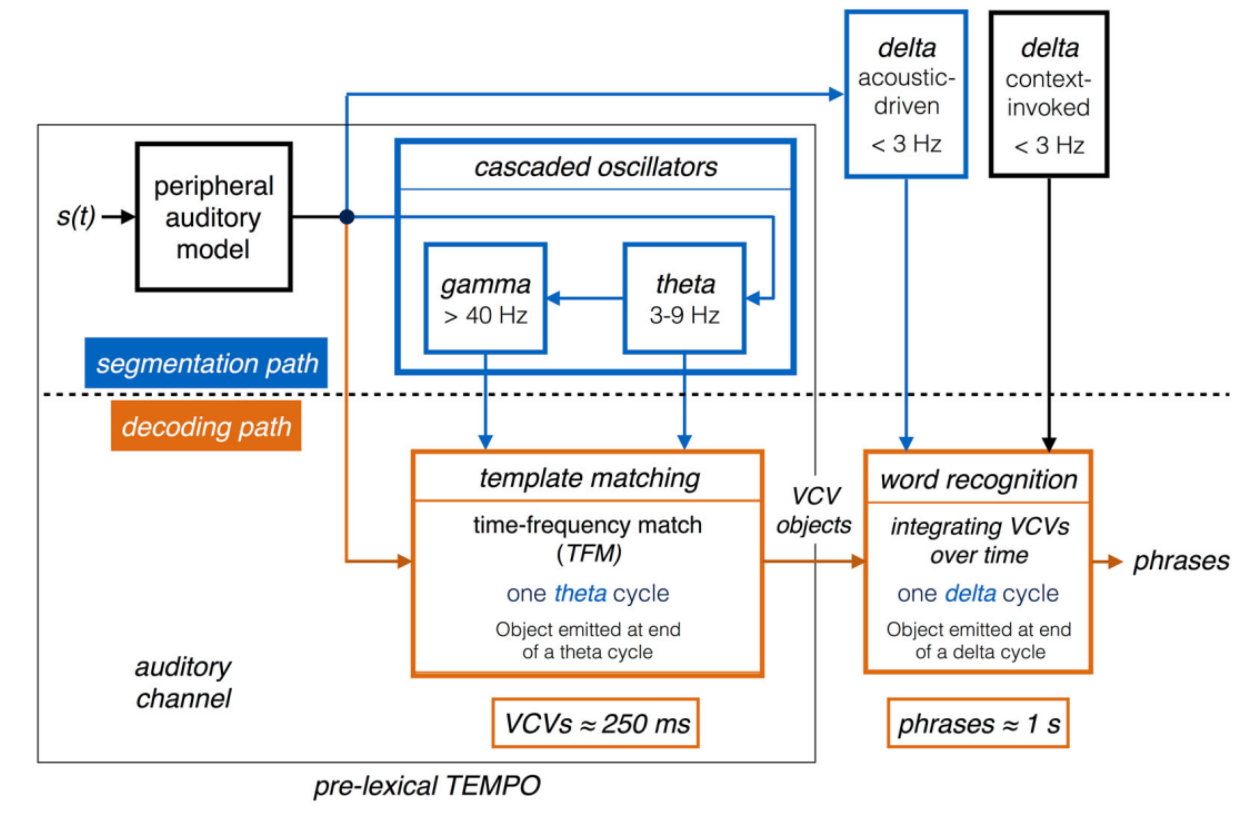}}
\end{minipage}
\caption{\textbf{Schematic of the TEMPO model.} This model comprises PAM \cite{ghitza1986auditory} for frequency decomposition, theta module for syllable segmentation, and gamma module for phoneme encoding. Delta module for control of word recognition process. Theta-gamma module is supposed to be implemented as the TFM model \cite{shamir2009representation}. Adapted from \cite{ghitza2017acoustic}.}
\label{fig:fig12}
\end{figure}

The TEMPO model \cite{ghitza2009possible} proposes to combine theta and gamma rhythms in order to explicitly account for the relationship between multi-rhythmic activity and speech perception \cite{ghitza2009possible}. Conceptually, the speech processing in the model can be divided into two components: segmentation and decoding. Speech segmentation occurs through the synchronization of theta, beta, and gamma rhythms. Phoneme decoding is accomplished by comparing the neural activity with patterns stored in memory (e.g., identified cell activity, as was done in models like TMC \cite{ghitza2007towards} or TFM \cite{shamir2009representation}). In later versions of the TEMPO model \cite{ghitza2017acoustic}, a delta rhythm (below 3 Hz) was added, corresponding to phrasal and lexical units of speech, which are temporal intervals from 400 to 4000 milliseconds (Fig.~\ref{fig:fig12}). The TEMPO model has been put forward to account for intriguing experimental findings \cite{ghitza2009possible}. In the experiment, speech was compressed/accelerated by a factor of 3 and thus rendered unintelligible. Accelerated speech is then segmented into 40 ms chunks. The key finding is that speech intelligibility is recovered when silences are inserted between the 40-ms chunks, such that each speech segment occurs at a normal syllabic rate . This experiment is fundamental because it shows that the speech intake capacity is not limited by the absolute information rate but by the information packaging. 

TEMPO lays out a blueprint that hypothesizes the principles by which multiple cascades of oscillations should result in invariant speech parsing. The TFM implements a reduced version of this idea (\cite{shamir2009representation}, where a neural pyramidal-interneuronal gamma-spiking network can parse artificial "saw-tooth" auditory signals (see Section 1.4). In subsequent work \cite{ghitza2011linking}, an extended version of this model is proposed conceptually with a hypothesized phenomenological mechanism, laying out the principles by which cascaded oscillators could enable the parsing of syllables in compressed and packaged speech. In the next TEMPO concept extension, \cite{ghitza2017acoustic} delta rhythm entrainment is proposed as a mechanism for the chunking of phrase candidates.

The Tempo model offers an interesting global conceptual framework but does not explicitly address the neural mechanism of theta synchronization with speech envelopes or the interaction between theta and gamma rhythms (see next section 3.2), nor was it explicitly implemented and tested on real speech signals. 

\subsubsection{The PING-PINTH model}

\begin{figure}[h]
\begin{minipage}[h]{1\linewidth}
\center{\includegraphics[width=1\linewidth]{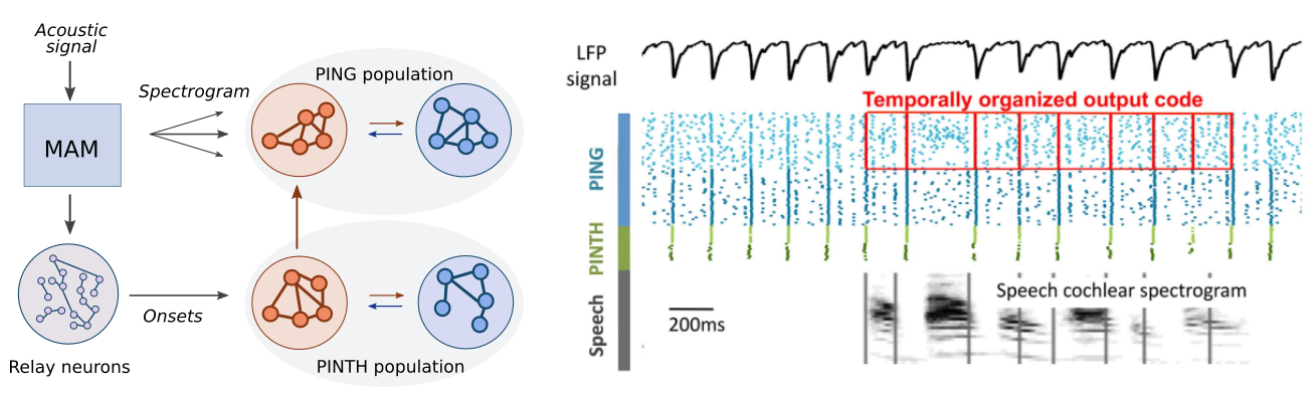}}
\end{minipage}
\caption{\textbf{Scheme of the PING-PINTH model.} Acoustic signal passes through a bandpass filter that is implemented using \cite{chi2005multiresolution}. Onset times are learned in the network of relay neurons. The PINTH network generates a theta rhythm that is modulated by the onset signal and segments speech into syllables. The PING network receives signals. about syllable boundaries and spectrograms. The activity of pyramidal neurons in the PING population encodes phonemes. (B) Network activity at rest and during speech perception. The rasterplot shows spikes from neurons: theta-inhibitory (dark green), theta-excitatory (light green), gamma-inhibitory (dark blue), and gamma-excitatory (light blue). The simulated local field potential (LFP) is at the top, with the auditory spectrogram at the bottom. Ge spikes represent the network's output. Adapted from \cite{hyafil2015speech}.}
\label{fig:fig13}
\end{figure}

The PING-PINTH model \cite{hyafil2015speech} has precisely been designed to explore the theta-gamma interactions using a biologically-based implementation with spiking leaky-integrate-and-fire neurons. It consists of two key neural network circuits; one generates an endogenous gamma rhythm through the classical Pyramidal-Interneuronal-Gamma mechanism (PING), and the second network generates the theta oscillations through a proposed pyramidal-interneuron mechanism (PINTH) \cite{hyafil2015speech} (Fig.~\ref{fig:fig13}). In each of these populations, the corresponding rhythms arise through synaptic interaction of the excitatory and inhibitory neurons. While the PING population follows standard synaptic parameters for gamma \cite{jadi2014regulating}, in the PINTH the inhibitory synapses are assumed to be slow so as to produce an endogenous theta rhythm. The PINTH network is trained to find syllable boundaries in continuous acoustic signals by locking to the minima in the acoustic signal envelope and generating spikes in the theta rhythm range. Its performance in tracking the variability of the syllabic rhythm is very close to that of human listeners (Fig.~\ref{fig:fig12}, vertical bars on the speech spectrogram). Its strength and originality lie in that it can syllabify in real time \cite{bartlett2009syllabification, demberg2007language}. The theta rhythm is a crucial mechanism that coordinates the activity of the PING network, creating a unified temporal context for speech perception. The PING network, on the other hand, generates the gamma rhythm, which is used to encode phonemes. The gamma rhythm is modulated by the theta rhythm from the PINTH network, allowing for a more reliable and efficient processing of phonemes.

Therefore, the PING-PINTH model was the first full-fledged biophysically-based model that successfully implements interconnected theta and gamma rhythms for speech recognition. The model explicitly demonstrates that coupled theta and gamma spiking rhythms can implement speech encoding and decoding. In particular, the theta rhythm, which parses syllables, modulates the gamma rhythm based on the rhythm of the syllables, supporting speech recognition. At the same time, viewed at the algorithmic level of speech processing, the coupling of theta and gamma rhythms leads to improved speech understanding, specifically via syllable-based recognition. This model also demonstrates its applicability, serving as the basis for investigating various effects, such as the impact of transcranial alternating current stimulation (tACS) on brain activity \cite{kegler2021modelling} and potential neuromorphic engineering solutions \cite{hyafil2015neuromorphic}.
However, the main PING-PINTH weakness is that it does not describe mechanisms for learning representations and the compositional effects that arise when dealing with complex semantic aspects of speech.

\section*{Discussion}

Currently, the majority of models for speech perception are primarily engineered for practical applications in artificial intelligence and automatic speech recognition \cite{young2018recent, patel2021deep, samant2022framework, olujimi2023nlp}. However, the question of the biological mechanisms actively involved in speech processing in the human brain remains unresolved. These biological mechanisms could be key to understanding the fundamental principles of perception and could serve to improve ASR algorithms by, e.g., optimizing basic sampling rhythms, including parsimonious predictive mechanisms, providing real-time syllabification, etc. In this paper, we strived to provide a focused survey of models that address the biological brain mechanisms of various aspects of speech perception, pointing out their advantages and pitfalls.

Computational modeling plays a vital role in understanding early auditory processing, spanning from the peripheral auditory system to higher cognitive stages. In the initial stages, the signal undergoes frequency decomposition, represented by a frequency spectrogram, upon entering the primary auditory cortex. This process is detailed in \cite{ghitza1986auditory, chi2005multiresolution, zilany2009phenomenological}.

Subsequently, oscillations associated with linguistic elements are involved in carrying out speech-specific operations in the auditory/temporal cortex \cite{giraud2012cortical}. These operations largely rely on orchestrating the timing of information processing (parsing, chunking, multiscale interactions). While gamma rhythms are associated with phoneme encoding \cite{mesgarani2014phonetic, tang2017intonational}, the theta rhythm can perform segmentation of speech into syllables \cite{gross2013speech, aiken2008human}, and the delta rhythm hypothetically governs word and phrase formation \cite{ghitza2017acoustic, rimmele2021acoustically}. These rhythm-based concepts were implemented in different computational models, offering possible approaches for mathematical understanding of the role of oscillations in the auditory cortex.

\subsection*{Gamma-scale phonemic information encoding}

Phoneme encoding can be implemented in biologically-based models in several ways. The TMC \cite{ghitza2007towards} implemented the phoneme representation encoded by the gamma rhythm in the pattern of spiking in neural networks. The primary problem in this model was that it did not implement temporal windowing for the signal identification and hence was not robust to speech rate variations and noise. 

This problem was subsequently addressed by the TFM \cite{shamir2009representation}, where a lower frequency rhythm is synchronized with the onset times of the input, segmenting the continuous acoustic signal into chunks for efficient signal processing. On the other hand, the Hopfield-Brody model \cite{hopfield2000moment, hopfield2001moment} and the WCO-TS model \cite{zavaglia2012dynamical} offer a mechanism of temporal sequence coding based on neuronal synchronization, where speech-feature-dependent synchronization required spike rate adaptation in individual neurons and detailed synaptic balance rather than synaptic plasticity as in WCO-TS. 

The TMC model \cite{ghitza2007towards} shows how a labeled line code could be implemented to identify phonemes, and the PING-PINTH \cite{hyafil2015speech} proposes a biologically plausible mechanism for gamma encoding of phonemes directly from the spectrogram, within theta-syllables. While two main mechanistic models have been proposed for the gamma rhythm, the interneuronal gamma (ING) and the pyramidal-interneuronal gamma (PING), experimental data suggests that it is PING that is most likely to occur in the cortex \cite{tiesinga2009cortical}. 

In fact, cortical gamma refers to a very broad frequency range (30 to over 100 Hz), which obviously cannot underpin the same operations. The higher the frequency, the closer it is to spiking activity, which is essentially a discrete phenomenon that has nothing to do with an oscillation per se. This variety of gammas, notably its bursty nature, is not reflected in the model. Thus, an open question to resolve is: how can phoneme coding be implemented with such bursty gamma, and might gamma bursts be involved in active coding of phonemes. This question can be addressed by a comprehensive exploration of dynamical activity in neurons during phoneme encoding. For example, it can be done by the low-rank RNN approach \cite{jazayeri2021interpreting} that allows to find low-dimensional representations of dynamical manifolds for the neural network activity during speech processing \cite{sussillo2013opening}. This approach can help to identify and distinguish different features coded in the neuronal activity and associate them with different stages of processing. Then, by the transition from training on experimental data RNN to more biological network representation—SNN, the gamma module of the PING-PINTH model could be modified, improving biological plausibility.

In the above models (with the exception of the PING-PINTH, see below), borders of relevant signals were explicitly implemented by hard-coded occurrence inputs. Yet speech-dependent coding of relevant borders is instrumental for processing by the auditory cortex of syllabic information. To model this process, we may consider experimental observations of the theta rhythm during speech perception in the auditory cortex, which appear to synchronize flexibly with the speech envelope \cite{aiken2008human, gross2013speech}. This rhythm can be described in several ways: at the level of individual neurons with the MI cell model \cite{pittman2021differential}, at the firing-rate network level with the AFO model \cite{doelling2023adaptive}, or generated by an interaction of excitatory and inhibitory spiking neurons as in the PING-PINTH model. Based on our reading of the HB and WCO-TS models, we may suggest that in these models, a theta time-scale may arise from entrained decay for syllable integration rather than from an endogenous generator. 

In this context, several issues need to be addressed. First, how to validate the model assumptions about the nature of the theta rhythm. For instance, one can test whether individual cells in the auditory cortex can flexibly adapt to pseudorhythmic speech-like signals and how this is influenced by specific ionic currents \cite{pittman2021differential}. Validating the AFO model causally would require identifying the source of the theta rhythm that generates the dynamic prior and how this endogenous rhythm is modulated by the auditory inputs. One way to measure the response of an endogenous theta rhythm to inputs is to identify the phase response curve to pulsatile stimulation (e.g., non-invasively by TMS/TdCS 
 or in subjects with implanted electrodes). 

According to the oscillatory hypothesis \cite{giraud2012cortical}, the rhythmic structure of cortical activity in the speech-related streams may act as a constraining factor on both the hierarchy structure of speech and the hierarchical rhythmic structure of speech processing in the brain, which itself reflects motor and biological constraints that are almost universal \cite{piette2024animal}. Along with findings from computational modeling, these hypotheses lead to a mechanistic hierarchical architecture for speech processing in the auditory cortex. A blueprint for how integrated theta-gamma code of syllables could be structured was proposed in TEMPO \cite{ghitza2007towards}. The first full biophysically-based implementation of this idea was presented as the PING-PINTH model \cite{hyafil2015speech}. This model stands out for its more realistic division into two interacting populations according to their functions and temporal dynamic scales: theta syllable segmentation by the PINTH population and gamma phoneme encoding by the PING one. In the model, the amplitude of the PING is coupled to the phase of the PINTH, and such phase-amplitude coupling (PAC) between rhythms has been shown to enhance performance in speech recognition tasks. 

\subsubsection*{Theta oscillation for tracking speech pseudorhythmicity}

\begin{figure}[h]
\begin{minipage}[h]{1\linewidth}
\center{\includegraphics[width=1\linewidth]{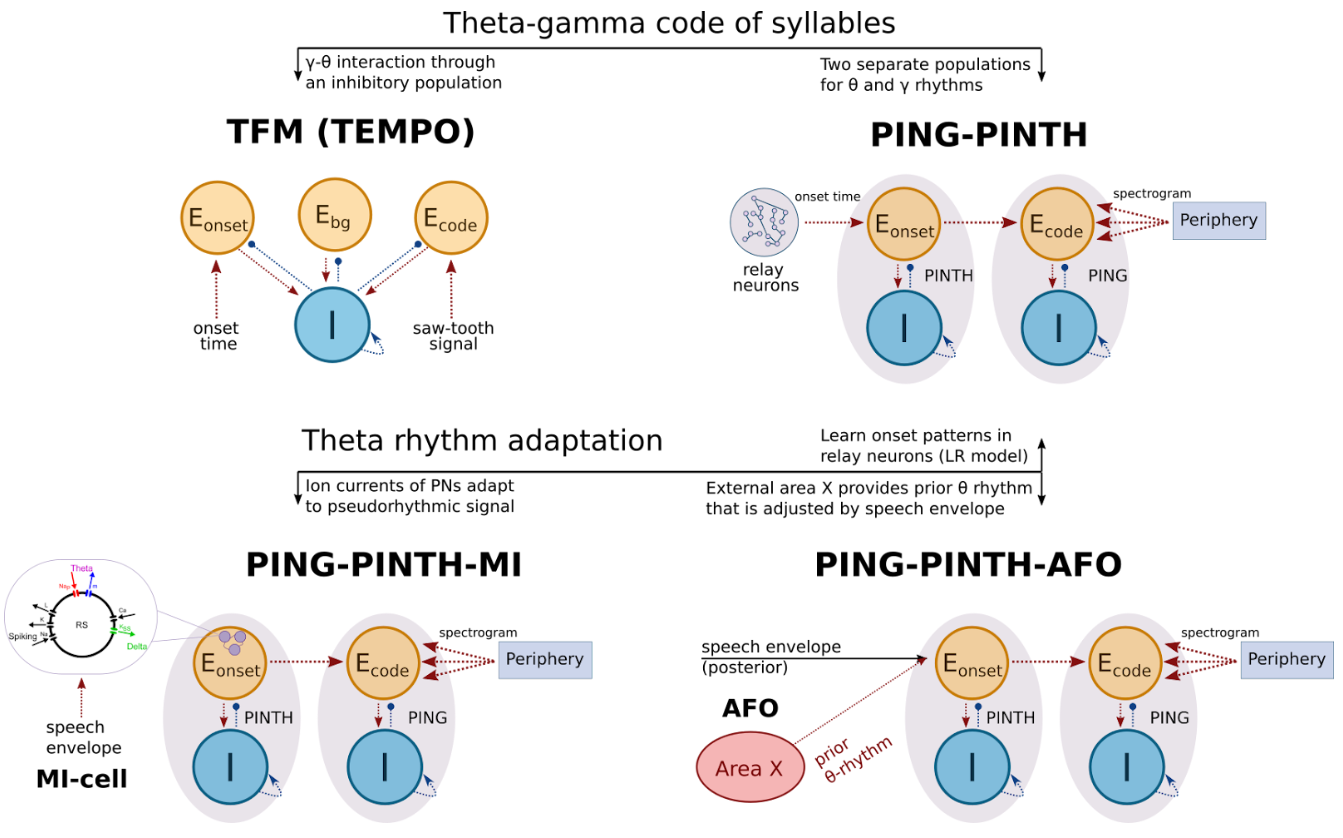}}
\end{minipage}
\caption{\textbf{Models for theta-gamma code of speech and possible future extensions.} Top panel: The TFM (TEMPO) and the PING-PINTH models represent two mechanisms for the theta-gamma code of lexical units (syllables and phonemes). Bottom panel: The PING-PINTH model can be improved by adding a more mechanistically plausible module for theta rhythm adaptation to speech envelope. It can be achieved by modification of the excitatory population in the PINTH module. Namely, one can replace LIF neurons with MI cells (left). Another option is the implementation of the AFO model instead of the relay neurons module (right).}
\label{fig:fig14}
\end{figure}

In the PING-PINTH model, the variability in syllable rate depending on speaker identity or emotional state is dealt with by including a weak theta oscillator that is nearly silent at rest and strongly engaged by the speech contours (syllable boundaries). This tracking is achieved by inputs from “relay”- neurons, implemented as a pre-trained filter. This relay-filter, unlike the other aspects of the model, lacks biophysical plausibility and detail. To implement a more realistic locking mechanism, the AFO model proposes an endogenously generated theta rhythm input that modulates the WC oscillator network and works as a temporal (predictive) template. The WC oscillator network adapts to an external acoustic signal to predict syllable onset times \cite{doelling2022adaptive}. Assaneo et al. proposed that the motor cortex could serve as the source of this prior rhythm, according \cite{assaneo2021speaking}. Alternative hypotheses may be that theta has a broader and more primitive origin than speech, whether motor or sensory, and is somehow related to hippocampal theta \cite{schroeder2009low, canolty2010functional, tort2009theta} or to multisensory active sensing (eye movements, sniffing, \cite{lakatos2007neuronal, chandrasekaran2009natural, arnal2009dual}. Finally, a concrete mechanism of theta rhythm emergence based on ion current dynamics was demonstrated in the MI cell model \cite{pittman2021differential}. The latter two models \cite{doelling2022adaptive, pittman2021differential} shed light on the mechanisms operating theta locking at both the population and individual neuron levels, and that can be used for a syllable parsing. Both models might be considered for relay-module replacement in the PING-PINTH model (Fig.~\ref{fig:fig14}).

In the PINTH population, excitatory neurons are described by the MI cell model with corresponding currents, and in the latter, there is a theta generator, and pyramidal neurons in PINTH adapt their frequency to the speech envelope. The AFO model avoids the requirement that the theta rhythm arises from within the auditory cortex. However, the motor cortex hypothesis is one among several others.

The models surveyed in this review demonstrate mechanisms that exhibit invariance to speech rates, voices, and acoustic variations. Invariance to speech rate is supported at the syllable segmentation level using theta rhythms, while invariance to acoustic pitch variations in voice is grasped by flexible gamma coding. These invariance properties lend credibility to the implemented mechanisms and formulated hypotheses. Yet some phenomena are not explained by these models, notably speech restoration/completion for interrupted and temporally segmented speech \cite{miller1950intelligibility, huggins1975temporally}, which require semantic language components and contextual influence. The early phase of segmenting speech into words and phrases, facilitated by the delta rhythm, is discussed for the TEMPO model \cite{ghitza2017acoustic}, but it doesn’t propose a specific neuronal implementation. Thus, at this stage, biologically-based models do not yet reach the semantic, contextual, and syntactic levels due in part to the limited exploration of this aspect using high-resolution neurophysiology (e.g., i-EEG, laminar-resolved data). 

\subsubsection*{The role of delta and beta rhythms in top-down contextual support}

The information processing hierarchy in the brain involves two directions of information flow \cite{gilbert2007brain, kveraga2007top, rabinovich2012information}. The bottom-up flow begins with the encoding of simple acoustic units, such as phonemes and syllables, and progresses to more complex semantic units like words and phrases. Simultaneously, the top-down flow shapes the contextual influence of semantics. For example, understanding the context provided by words and phrases helps infer details about their constituent syllables and phonemes. However, in a feedforward approach, phonemes and syllables function as auxiliary elements without direct semantic meaning; their primary role is to activate the relevant words and phrases. Experimental evidence suggests that the delta rhythm is associated with top-down information flow, and one of its hypothesized roles is to govern contextual support during speech processing \cite{park2020predictive, ten2022neural, kaufeld2020linguistic}.

Thus, phonemes and syllables represent the basic units of speech, processed at a lower level and passed bottom-up. Words and phrases, in contrast, contain semantic meaning. In unclear contexts, they are gradually assembled with syllables and phonemes, progressively reducing uncertainty until the correct word or phrase is identified. This process is related to the predictive coding concept \cite{donhauser2020two, friston2008variational} and involves explicit prediction of words and phrases, followed by inference and information updating as needed. The output, facilitated by top-down information flow, is then used to minimize uncertainty in subsequent syllables and phonemes.

Models that integrate the predictive coding approach with oscillatory mechanisms have already been developed. For instance, the Precoss model \cite{hovsepyan2020combining} incorporates the coupled theta-gamma rhythm for syllable and phoneme processing, and its extension, Precoss-beta \cite{hovsepyan2023rhythmic}, demonstrates how the beta rhythm orchestrates the prediction-correction process during recognition. The BRyBI model \cite{dogonasheva2024brain} goes beyond syllable coding to include word and phrase levels, demonstrating how the delta rhythm governs the integration time windows for contextual information. Another notable model within the predictive coding framework, the Su-model \cite{su2023deep}, highlights the significance of considering the contextual information flow in speech processing, particularly at the levels of semantic representations and syntax.

Although these models employ oscillatory mechanisms, they describe rhythms more phenomenologically rather than using neural elements, which places them outside the primary focus of this review. Nevertheless, the future of neuro-oscillatory models within the predictive coding framework represents a promising direction for more comprehensive modeling of speech processing.

While these biologically-based models are essential to grasping the details of speech computations in the human brain, they are often complex and demand significant computational resources for implementation and analysis, imposing drastic trade-offs in choice of parameters. Although this complexity may pose challenges in extending their applicability, optimizing and simplifying biologically-based models could make them more accessible and suitable for real-world applications. 

\section*{Conclusion}

Despite their complexity and computational requirements, biologically-based models are useful to explain invariance to different conditions, such as speech rates and speaker identity. They enable researchers to gain insights into how real neurons and networks can process and perceive speech. Biologically-based models can explain how the brain segments speech into smaller units while simultaneously integrating information within these segments—a crucial property for successful speech recognition. However, they may face limitations in processing interrupted and heavily distorted speech, where semantics and context become crucial factors. These aspects will have to be taken into account and modeled with equal biological plausibility.

Such demanding and realistic computational models promise to contribute not only to understanding the fundamental mechanisms of speech perception but also to the development of robust speech recognition systems resilient to various real-time conditions. Further integration of biological concepts and artificial intelligence will provide new insights into the research of speech perception and its applications.

\section*{BOX}

Before reaching the cortex, the speech signal is preprocessed by an exquisitely fine-tuned peripheral system that faithfully encodes the sound's spectro-temporal details. Models of speech processing by the cortex must thus be fed with signals coming from realistic models of the peripheral auditory system. Among those used in this field we can cite: the Peripheral Auditory Model (PAM) \cite{ghitza2007towards}, the Multiresolution Auditory Model (MAM) \cite{chi2005multiresolution}, and Power-Law Adaptation (PLA) \cite{zilany2009phenomenological}. These models have been designed to fit experimental data and implement mechanisms of nonlinear transformations and frequency filtering observed in the path of speech signal processing from the cochlea to the midbrain.

\begin{figure}[h]
\begin{minipage}[h]{1\linewidth}
\center{\includegraphics[width=1\linewidth]{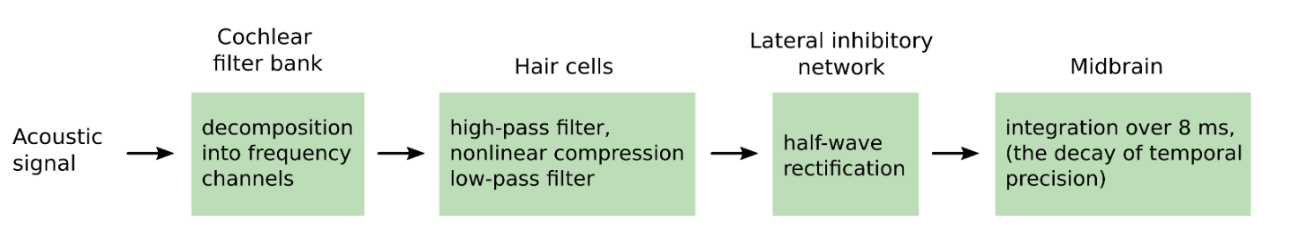}}
\end{minipage}
\caption{\textbf{Scheme of the Multiresolution Auditory Model.} The acoustic signal undergoes analysis through a set of cochlear-like filters. Each filter's output is then processed through a hair cell model, followed by a lateral inhibitory network. The final step involves rectification and integration, resulting in the auditory spectrogram. Adapted from \cite{chi2005multiresolution}.}
\label{fig:figBOX}
\end{figure}

For example, MAM \cite{chi2005multiresolution} decomposes the signal into 128 frequency channels using a cochlear filter bank, followed by nonlinear transformations in the inner hair cells, cochlear nucleus, and midbrain. This model enhances sound contrast and discrimination through lateral inhibition, allowing selective attention to relevant auditory stimuli. The final processing stage in the midbrain integrates these signals over short temporal intervals, reflecting the decay of temporal precision and reducing noise.

\section{Acknowledgements}
This work/article is an output of a research project implemented as part of the Basic Research Program at the National Research University Higher School of Economics (HSE University). This work has benefited from a French government grant managed by the Agence Nationale de la Recherche under the France 2030 program, reference ANR-23-IAHU-0003. This work was supported by a grant from Fondation Pour l’Audition (FPA) to Anne-Lise Giraud (FPA IDA11).
BSG was supported by CNRS, INSERM.

\bibliographystyle{unsrt}
\bibliography{main}

\end{document}